\PassOptionsToPackage{table,xcdraw,dvipsnames}{xcolor}

\documentclass{article}

\PassOptionsToPackage{numbers, compress}{natbib}
\usepackage[preprint]{template}

\usepackage[utf8]{inputenc} 
\usepackage[T1]{fontenc}    
\usepackage{hyperref}       
\hypersetup{
    colorlinks=true,
    linkcolor=BlueViolet,
    citecolor=teal
}
\usepackage{url}            
\usepackage{booktabs}       
\usepackage{amsfonts}       
\usepackage{nicefrac}       
\usepackage{microtype}      
\usepackage{xcolor}         

\usepackage{makecell}
\usepackage{amsthm}
\usepackage{bm}
\usepackage[table,xcdraw]{xcolor}
\usepackage{multicol}
\usepackage{colortbl}
\usepackage{amsmath}
\usepackage{cleveref}
\usepackage{multirow}
\usepackage{graphicx}
\usepackage{subfigure}
\usepackage{enumitem}
\usepackage{wrapfig}
\usepackage{bbding}
\usepackage{color}

\usepackage{algorithm}
\usepackage{listings}
\usepackage{natbib}
\usepackage{tikz}
\usepackage{amssymb}
\usepackage{amsmath}

\usepackage{etoolbox}
\makeatletter
\AfterEndEnvironment{algorithm}{\let\@algcomment\relax}
\AtEndEnvironment{algorithm}{\kern2pt\hrule\relax\vskip3pt\@algcomment}
\let\@algcomment\relax
\newcommand\algcomment[1]{\def\@algcomment{\footnotesize#1}}
\renewcommand\fs@ruled{\def\@fs@cfont{\bfseries}\let\@fs@capt\floatc@ruled
  \def\@fs@pre{\hrule height.8pt depth0pt \kern2pt}%
  \def\@fs@post{}%
  \def\@fs@mid{\kern2pt\hrule\kern2pt}%
  \let\@fs@iftopcapt\iftrue}
\makeatother

\definecolor{ForestGreen}{RGB}{34,139,34}

\newcommand{\ie}{\emph{i.e.},~}

\renewcommand{\paragraph}[1]{\medskip\noindent\textbf{#1.~}}

\newcommand{\shuk}[1]{\left\lVert#1\right\rVert}

\newcommand{\argmin}[1]{{\mathop{\arg\mathrm{min}}_{#1}\,}}

\newcommand{\bmf}{\bm{f}}

\newcommand{\bmr}{\bm{r}}
\newcommand{\bms}{\bm{s}}

\newcommand{\bmw}{\bm{w}}
\newcommand{\bmx}{\bm{x}}
\newcommand{\bmy}{\bm{y}}
\newcommand{\bmz}{\bm{z}}

\newcommand{\bmtheta}{\bm{\theta}}

\newcommand{\bmmu}{\bm{\mu}}

\newcommand{\bmsigma}{\bm{\sigma}}

\newcommand{\bmpsi}{\bm{\psi}}

\newcommand{\bmA}{\bm{A}}
\newcommand{\bmB}{\bm{B}}
\newcommand{\bmC}{\bm{C}}

\newcommand{\bmI}{\bm{I}}

\newcommand{\bbE}{\mathbb{E}}

\newcommand{\bbR}{\mathbb{R}}

\newcommand{\modelname}{MambaVC}

\title{\modelname{}: Learned Visual Compression with Selective State Spaces}

\author{
	Shiyu Qin\textsuperscript{\rm 1}, 
	Jinpeng Wang\textsuperscript{\rm 1}, 
	Yimin Zhou\textsuperscript{\rm 2}, 
	Bin Chen\textsuperscript{\rm 2,\rm 5,\rm 6}\thanks{Corresponding author}, 
        Tianci Luo \textsuperscript{\rm 2}, \\
	\textbf{Baoyi An\textsuperscript{\rm 3}},
        \textbf{Tao Dai\textsuperscript{\rm 4}},
        \textbf{Shutao Xia\textsuperscript{\rm 1}},
        \textbf{Yaowei Wang\textsuperscript{\rm 5}}\\
	\textsuperscript{\rm 1}Tsinghua Shenzhen International Graduate School, Tsinghua University.\\
	\textsuperscript{\rm 2}Harbin Institute of Technology, Shenzhen. \textsuperscript{\rm 3}Huawei Technologies Company Ltd.\\
        \textsuperscript{\rm 4}Shenzhen University. \textsuperscript{\rm 5}Peng Cheng Laboratory.\\
        \textsuperscript{\rm 6}Guangdong Provincial Key Laboratory of Novel Security Intelligence Technologies.\\
	{\tt\small \{qinsy23, wjp20\}@mails.tsinghua.edu.cn}; {\tt\small \{200110126, 210310217\}@stu.hit.edu.cn}; \\
        {\tt\small chenbin2021@hit.edu.cn}; {\tt\small anbaoyi@huawei.com}; {\tt\small daitao.edu@gmail.com};  \\
        {\tt\small xiast@sz.tsinghua.edu.cn}; {\tt\small wangyw@pcl.ac.cn};
}

\begin{document}

\maketitle
\vspace{-0.2cm}
\begin{abstract}
Learned visual compression is an important and active task in multimedia. 
Existing approaches have explored various CNN- and Transformer-based designs to model content distribution and eliminate redundancy, where balancing efficacy (\ie rate-distortion trade-off) and efficiency remains a challenge. 
Recently, state-space models (SSMs) have shown promise due to their long-range modeling capacity and efficiency. 
Inspired by this, we take the first step to explore SSMs for visual compression. 
We introduce \modelname{}, a simple, strong and efficient compression network based on SSM. 
\modelname{} develops a visual state space (VSS) block with a 2D selective scanning (2DSS) module as the nonlinear activation function after each downsampling, which helps to capture informative global contexts and enhances compression. 
On compression benchmark datasets, \modelname{} achieves superior rate-distortion performance with lower computational and memory overheads. 
Specifically, it outperforms CNN and Transformer variants by 9.3\% and 15.6\% on Kodak, respectively, while reducing computation by 42\% and 24\%, and saving 12\% and 71\% of memory. 
\modelname{} shows even greater improvements with high-resolution images, highlighting its potential and scalability in real-world applications. 
We also provide a comprehensive comparison of different network designs, underscoring \modelname{}'s advantages.
Code is available at \url{https://github.com/QinSY123/2024-MambaVC}.
\end{abstract}
\vspace{-0.2cm}
\section{Introduction}
\label{sec: introduction}
Visual compression is a long-standing problem in multimedia processing. In the past few decades, classical standards~\cite{bellard2018bpg, bross2021overview} have dominated for a long time. With the advent of deep neural architectures like CNNs~\cite{balle2016end,balle2018variational,cheng2020learned,duan2023lossy,he2022elic,wang2022neural} and Transformers~\cite{koyuncu2022contextformer, qian2021entroformer,zhu2021transformer,zou2022devil}, learned compression methods have emerged and shown ever-improving performance, gaining increasing interest over traditional ones.


The core of visual compression is the neural network design to eliminate redundant information and capture content distribution, where it naturally presents a dilemma between rate-distortion optimization and model efficiency. 
While CNN-based methods~\cite{balle2016end,balle2018variational,cheng2020learned,duan2023lossy,he2022elic,wang2022neural} remain popular in many resource-limited scenarios thanks to the hardware-efficient convolution operators, 
their local receptive field~\cite{luo2016understanding} limits global context modeling capacity and thus restricts compression performance. 
In contrast, Transformer-based methods~\cite{koyuncu2022contextformer, qian2021entroformer,zhu2021transformer,zou2022devil} excel in the global perception with attention mechanisms and thereby benefit redundancy reduction. 
However, their quadratic complexity in computation and memory raises efficiency concerns. 
Although some hybrid approaches like TCM~\cite{liu2023learned} combine CNNs and Transformers to balance compression efficacy and efficiency, it is not a sustainable direction for further development. 
Unlike prior work, we are committed to exploring promising solutions beyond engineering trade-offs toward this issue. 

Recently, state space models (SSMs)~\cite{gu2023mamba,mehta2023long,wang2023selective}, particularly the structured variants (S4)~\cite{gu2021efficiently}, have been extensively studied. Mamba~\cite{gu2023mamba} stands out as a representative work, whose data-dependent selective mechanism enhances critical information extraction while eliminating irrelevant noise from the input. 
This hints that Mamba-based models can effectively gather global context and thus enjoy advantages for compression. 
Furthermore, Mamba integrates structured reparameterization tricks and utilizes a hardware-efficient parallel scanning algorithm, assuring faster training and inference on GPUs.
These compelling features inspire us to investigate Mamba's potential for visual compression.

\begin{figure}[t]
	\centering
	\subfigure[BD-rate (lower is better) vs computational complexity and memory overhead (circle area) on Kodak~\cite{franzen1999kodak}.\label{fig:BDrate_MAC_size}] 
    {\includegraphics[width=.49\textwidth]{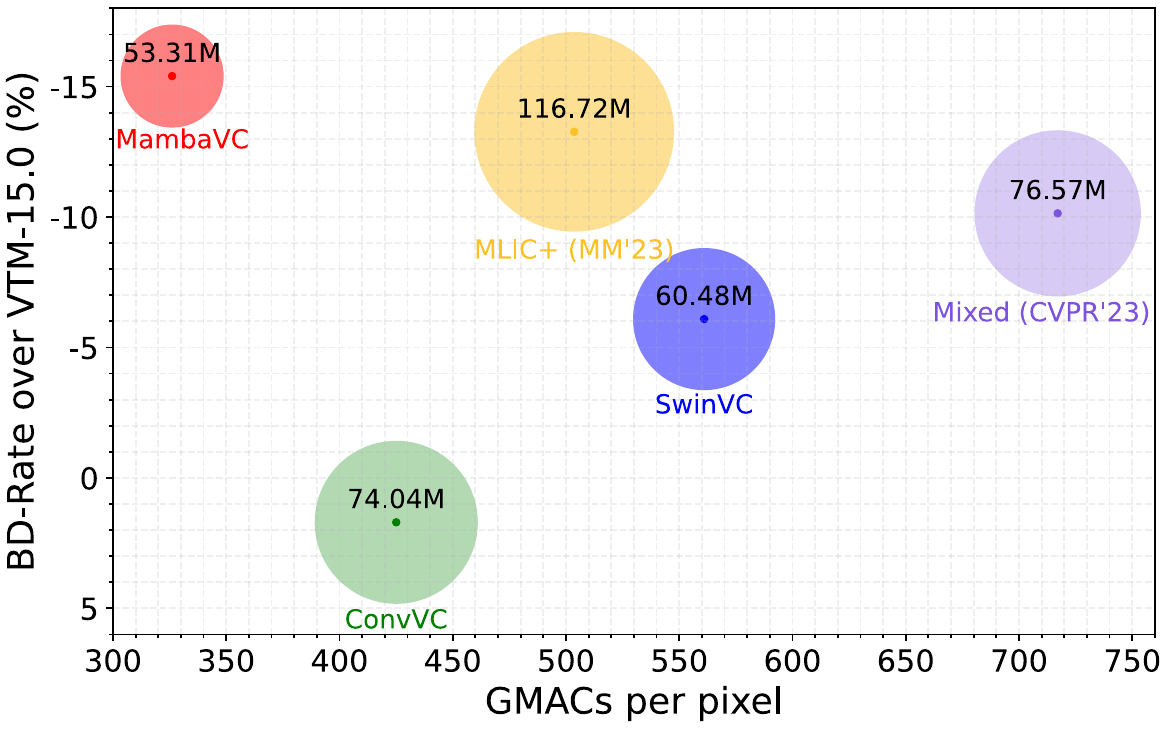}}
    \hspace{0.006\textwidth}
	\subfigure[BD-rate of \modelname{} over variants across different image resolutions on UHD~\cite{zhang2021benchmarking}.\label{fig:image_resolution}]     
    {\includegraphics[width=.49\textwidth]{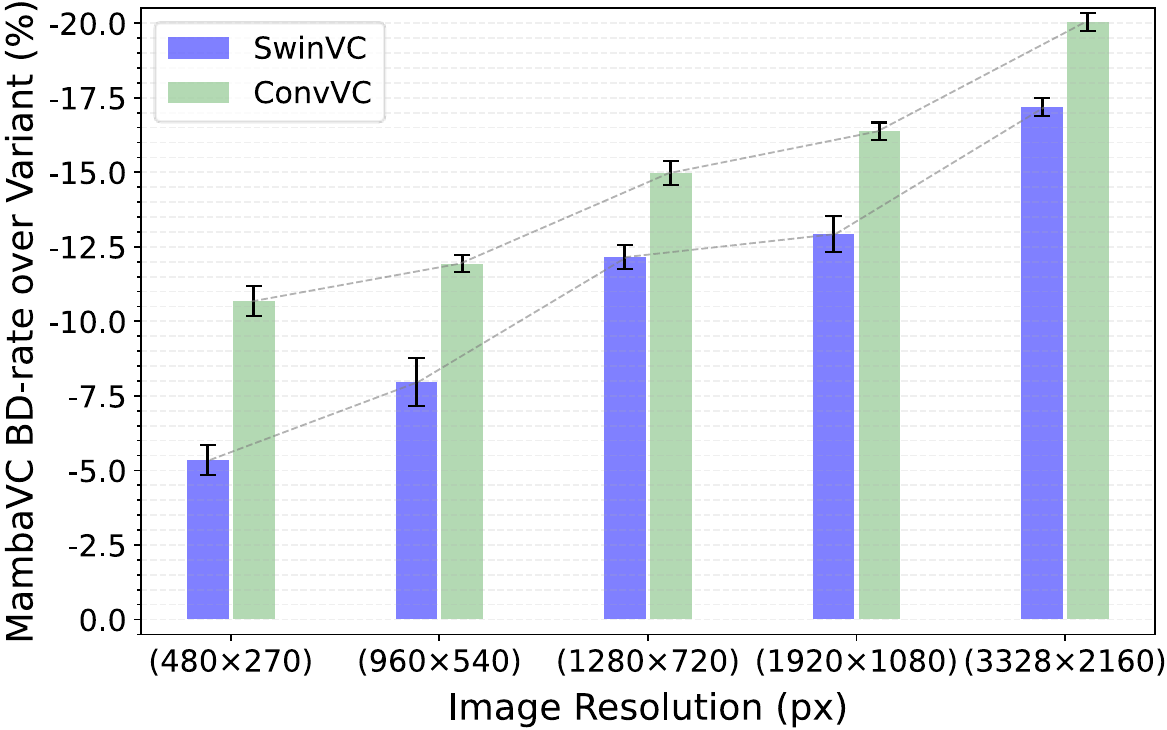}}
    \vspace{-0.2cm}
	\caption{(a) \modelname{} achieves the best BD rate with the least computation and memory overhead.
 See \Cref{Variant Visual Compression Performance} and \Cref{Computational complexity} for more details. 
 (b) The improvements of \modelname{} over other designs becomes more pronounced with increasing resolutions.}
	\label{Fig_intro}
    \vspace{-0.4cm}
\end{figure}

In this paper, we introduce \modelname{}, a simple, strong and efficient visual compression network with selective state spaces. 
Inspired by \citet{liu2024vmamba}, we design a \emph{visual state space} (VSS) block as the nonlinear activation function after each downsampling in the neural compression network, which integrates a specialized \emph{2D selective scanning} (2DSS) mechanism for spatial modeling.  
The 2DSS performs selective scanning along 4 pre-defined traverse paths in parallel, which helps to capture comprehensive global contexts and facilitates effective and efficient compression.

We conduct extensive experiments on image and video benchmark datasets. 
Without the bells and whistles, \modelname{} achieves a superior rate-distortion trade-off with lower computational and memory overheads compared to CNN- and Transformer-based counterparts, some as demonstrated in \Cref{fig:BDrate_MAC_size}. 
More encouragingly, we show that \modelname{} exhibits even stronger performance on high-resolution image compression, as shown in \Cref{fig:image_resolution}. 
These favorable results are consistent with SSM's efficient long-range modeling capacity, shedding light on its potential in many important yet challenging applications, such as compressing high-definition medical images and transmitting high-resolution satellite imagery. 
We also compare and analyze different designs from various aspects, including spatial redundancy, effective receptive field, and information loss in the compression process, to facilitate a comprehensive understanding of \modelname{}'s efficacy.

In summary, our contributions are as follows:
\setlist{nolistsep}
\begin{itemize}[leftmargin=1.5em]
\item We develop \modelname{}, the first visual compression network with selective state spaces. The designed 2DSS improves global context modeling and helps effective and efficient compression. 
\item Extensive experiments on benchmark datasets show superior performance and competitive efficiency of \modelname{} on image and video compression. 
The strong results highlight a new promising direction of compression network design beyond CNNs and Transformers. 
\item We showcase \modelname{}'s particular effectiveness and scalability in high-resolution compression, prompting its potential in many important but challenging applications. 
\item We compare and analyze 
different network designs thoroughly, showing the \modelname{}'s advantages regarding various aspects to validate and understand its effectiveness.  

\end{itemize}


\vspace{-0.2cm}
\section{Related Works}
\label{sec:related_work}
\textbf{Learned Visual Compression} In the past decade, learned visual compression has demonstrated remarkable potential and made a significant impression. The prevailing methods can be categorized into CNN-based and Transformer-based approaches. Early works, such as CNNs with generalized divisive normalization (GDN) layers~\cite{balle2016end,balle2018variational, minnen2018joint}, achieved good performance in image compression. Attention mechanisms and residual blocks~\cite{cheng2020learned,zhang2019residual, zhou2019end} were integrated into the VAE architecture later. However, the limited receptive field constrained the further development of these models. With the explosion of Vision Transformers~\cite{dosovitskiy2020image, liu2021swin}, Transformer-based compression models~\cite{lu2022transformer, qian2021entroformer,zhu2021transformer,zou2022devil} have shown strong competitiveness. Yet, their substantial computational and storage demands are daunting. Recent efforts~\cite{liu2023learned} have attempted to combine the strengths of both approaches, but led to even increased computational complexity as shown in Figure~\ref{fig:BDrate_MAC_size}. The trade-off between model performance and efficiency remains a pressing issue that needs to be addressed.
 
\textbf{State Space Models} SSMs are recently proposed models combined with deep learning to capture the dynamics and dependencies of long-sequence data. LSSL~\cite{gu2021combining} first leverages linear state space equations for modeling sequence data. Later, the structured state-space sequence model (S4)~\cite{gu2021efficiently} employs a linear state space for contextualization and shows strong performance on various sequence modeling tasks, especially with lengthy sequences. Building on it, numerous ~\cite{fu2022hungry,mehta2023long,smith2022simplified} have been proposed, and Mamba~\cite{gu2023mamba} stands out with its data dependency and parallel scanning. Many works have consequently extended Mamba from Natural Language Processing (NLP) to the vision domain such as image classification~\cite{liu2024vmamba,zhu2024vision}, multimodal Learning~\cite{qiao2024vl} and others~\cite{chen2024video,guo2024mambair,ma2024u}. However, the application of the Mamba for visual compression remains unexplored. In this work, we explore how to transfer the success of Mamba to build effective and efficient compression models.

\vspace{-0.2cm}
\section{Method}
\label{sec:method}


\subsection{Preliminaries: State-State Models and Mamba}
State-space models (SSMs) map stimulation $x(t)\in\bbR^L$ to response $y(t)\in\bbR^L$ through a hidden state $h(t)\in\bbR^N$, where we define matrix $\bmA ^{N \times N}$ as the evolution mapping of the hidden state, matrices $\bmB ^{N \times 1}$ and $\bmC ^{1 \times N}$ as the input and readout mappings for the hidden state, respectively. Typically, we can formulate the process by linear ordinary differential equations (ODEs): 
\begin{equation}
\begin{aligned}
h'(t)&=\bmA h(t)+\bmB x(t), \\
y(t)&=\bmC h(t).
\label{ODE}
\end{aligned}
\end{equation}

Modern SSMs approximate this continuous-time ODE through
discretization. 
Concretely, they discretize the continuous parameters $\bmA$ and $\bmB$ by a timescale $\Delta$, using the zero-order hold trick:
\begin{gather}
\bar{\bmA}=\exp(\Delta\bmA), \\
\bar{\bmB}=(\Delta\bmA)^{-1}(\exp(\Delta\bmA)-\bmI)\cdot\Delta\bmB.
\end{gather}
Then the discretized version of \cref{ODE} is reformulated as follows:
\begin{equation}
\begin{aligned}
h_t&=\bar{\bmA}h_{t-1}+\bar{\bmB}x_t, \\
y_t&=\bmC h_t.
\label{linear recursive manner}
\end{aligned}
\end{equation}
Mamba~\cite{gu2023mamba} further incorporates data-dependence to $\Delta$, $\bmB$ and $\bmC$, enabling an input-aware selective mechanism for better state-space modeling. 
While the recurrent nature restricts the fully parallel capacity, Mamba ingeniously implements structural reparameterization tricks and the hardware-efficient parallel scanning algorithm to compensate for the overall efficiency. 

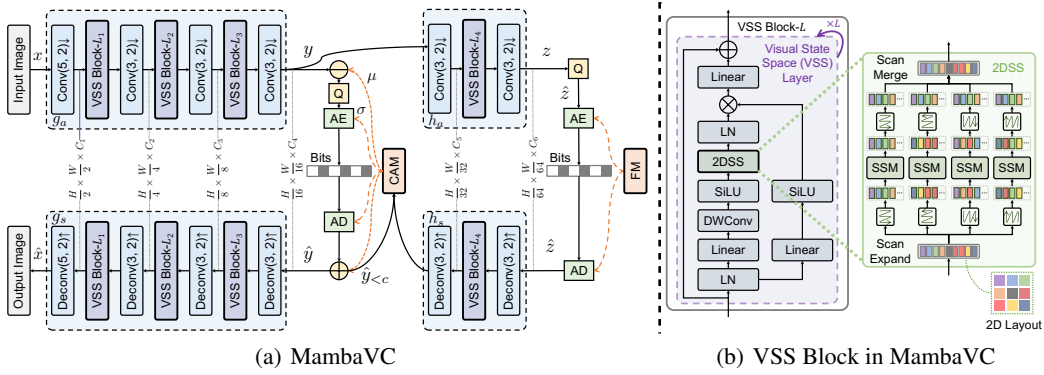
\begin{figure}[t]
	\centering
	\subfigure[\modelname{}\label{MambaVC}] 
    {\includegraphics[width=.6205\textwidth]{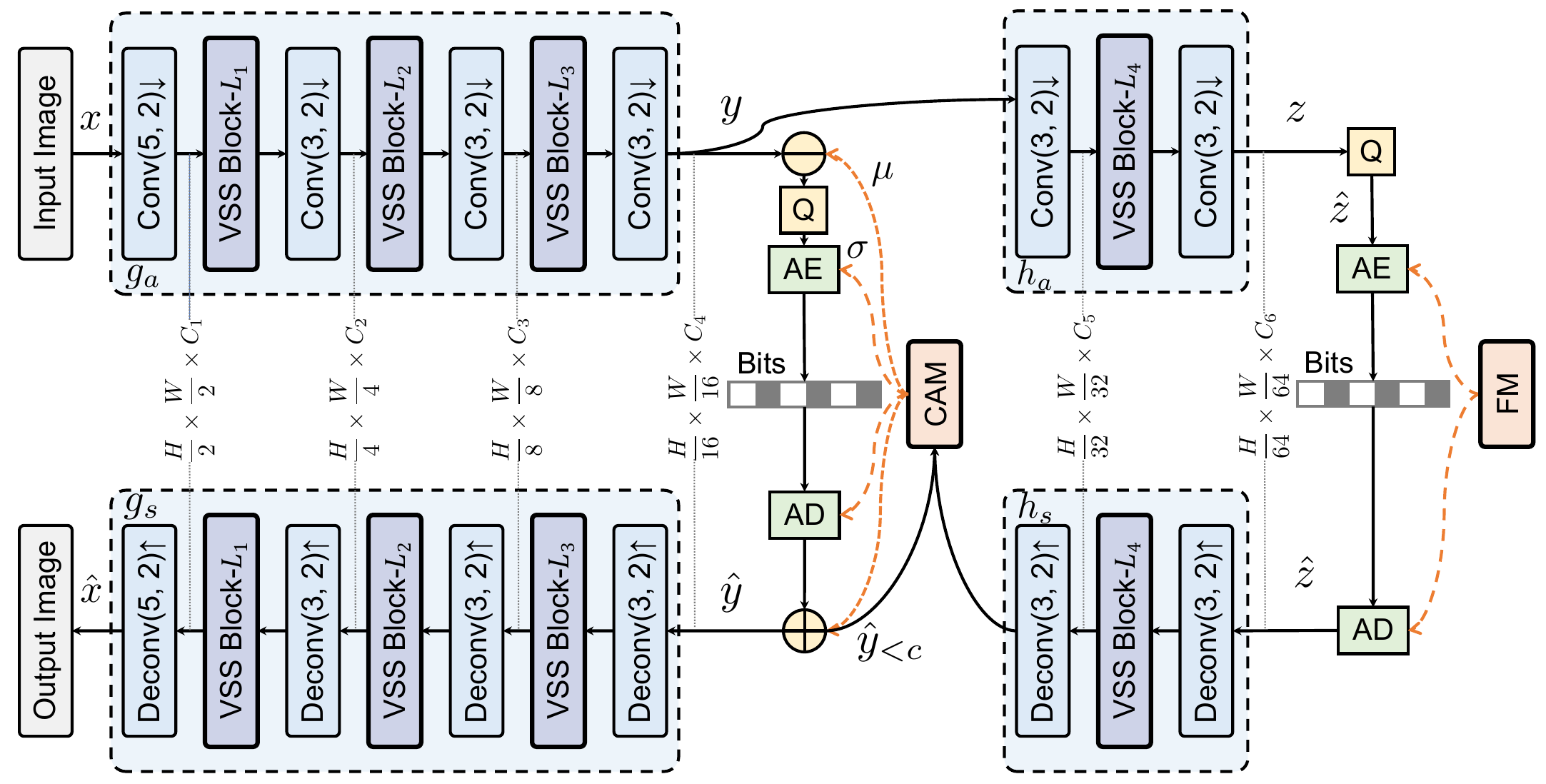}}
    \begin{tikzpicture}
        \draw[line width=1.2pt, dash pattern=on 2.5pt off 2.5pt, dash phase=2pt] (0,1.5) -- (0,6);
    \end{tikzpicture}%
	\subfigure[VSS Block in \modelname{}\label{VSSB}]     
    {\includegraphics[width=.370\textwidth]{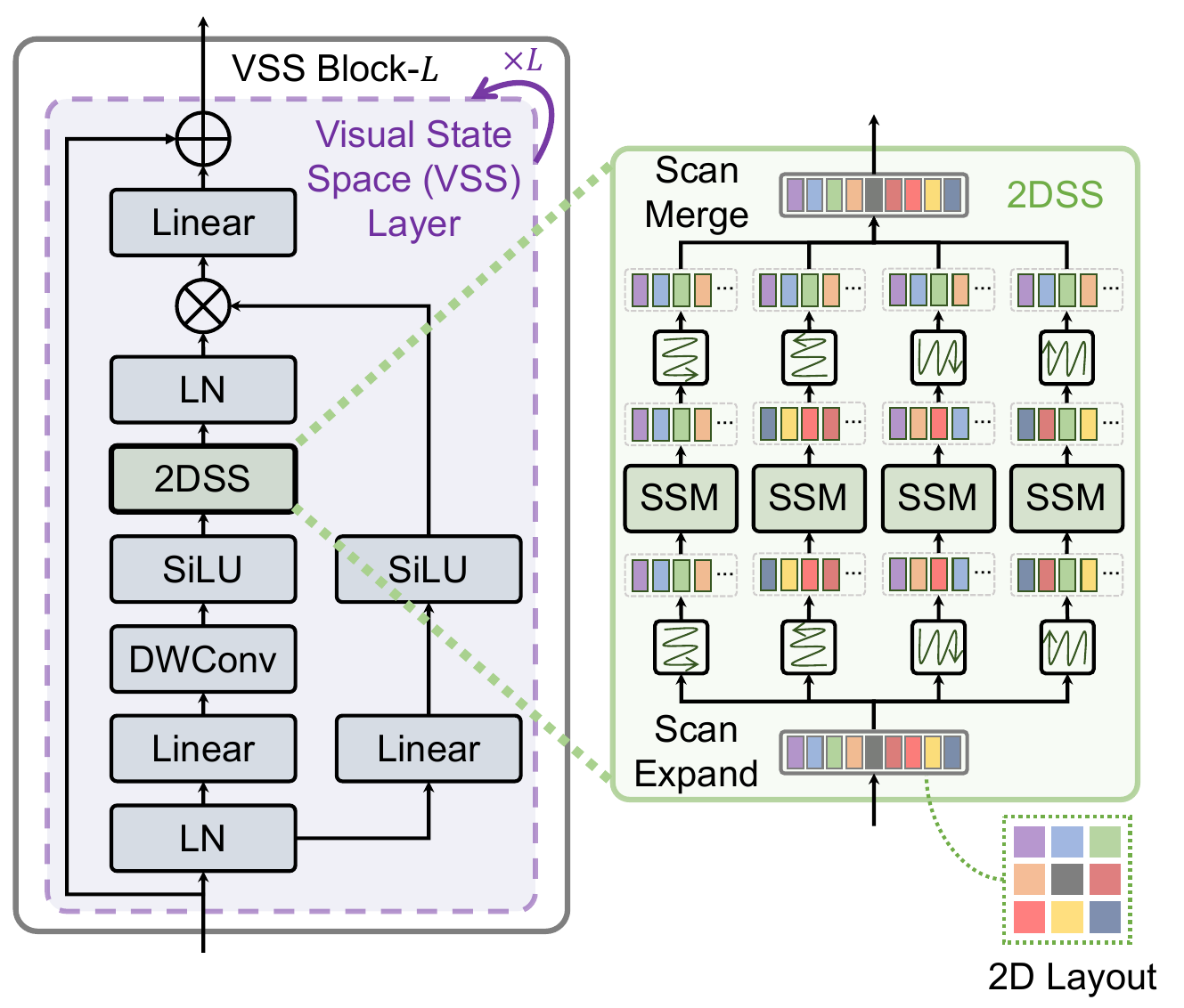}}
	\caption{(a) Overview of \modelname{}. \textbf{CAM} is channel-wise auto-regressive entropy model~\cite{liu2023learned}. \textbf{FM} is factorized entropy model. \textbf{$\text{Conv}(N, 2)\downarrow$} and \textbf{$\text{Deconv}(N, 2)\uparrow$} represent strided down convolution and strided up convolution with $N{\times}N$ filters, respectively. 
    (b) A VSS block consists of several layers. Each layer includes a 2DSS module, which performs selective scans in 4 parallel patterns.}
    \vspace{-0.3cm}
	\label{backbone}
\end{figure}
\subsection{The proposed \modelname{}}
\label{The proposed \modelname{}}
\subsubsection{Overview}
We illustrate the architecture of \modelname{} in \Cref{MambaVC}. 
Given an image $\bmx$, we first obtain the latent $\bmy$ and hyper latent $\bmz$ using the encoder $g_a$ and the hyper encoder $h_a$, respectively:
\begin{gather}
\bmy = g_a(\bmx; \bmtheta_{g_a}), \\
\bmz = h_a(\bmy; \bmtheta_{h_a}).
\end{gather}
Then, the quantized hyper latent $\hat{\bmz}=Q(\bmz)$ is entropy coded for rate $R(\hat{\bmz})=\bbE[-\text{log}_2(p_{\bmz | \bmpsi} \left( \hat{\bmz} \mid \bmpsi \right))]$, where $p_{\bmz | \bmpsi} \left( \hat{\bmz} \mid \bmpsi \right)=\Pi_{j} \left( p_{\bmz_j|\bmpsi}(\bmpsi) \ast \mathcal{U}\left( -\frac{1}{2}, \frac{1}{2} \right) (\hat{\bmz}_j) \right)
$, with a learned factorized prior $\bmpsi$. 

At the decoder side, we first use a hyper decoder $h_s$ to obtain the initial mean and variance: 
\begin{gather}
(\bm{\tilde{\mu}}, \bm{\tilde{\sigma}})=h_s(\hat{\bmz}; \bmtheta_{h_s}).
\end{gather}
Then we divide the latent $\bmy$ to $S$ slices $\bmy_0, \bmy_1, \cdots, \bmy_{S-1}$ and compute slice-wise information by:
\begin{gather}
\bmr_i, (\bmmu_i,\bmsigma_i) = e_i(\bm{\tilde{\mu}}, \bm{\tilde{\sigma}}, \bar{\bmy}_{<i}, \bmy_i;\bmtheta_{e_i}), \\
\bar{\bmy}_i = \bmr_i + \hat{\bmy}_i = \bmr_i + Q(\bmy_i-\bmmu_i)+\bmmu_i,
\end{gather}
where $e_i$ represents the $i$-th network in the channel-wise auto-regressive entropy model (CAM)~\cite{liu2023learned}, $i=0,1,\cdots,S-1$. 
We concatenate the slice-wise estimated distribution parameters and obtain the holistic $\bmmu$ and $\bmsigma$. We compute $R(\hat{\bmy})=\bbE[-\text{log}_2(p_{\hat{\bmy} | \hat{\bmz}} \left( \hat{\bmy} \mid \hat{\bmz} \right))]$ with $p_{\hat{\bmy} | \hat{\bmz}} \left( \hat{\bmy} \mid \hat{\bmz} \right)\sim\mathcal{N}(\bmmu,\bmsigma^2)$.

Next, we use the decoder $g_s$ to reconstruct image from the quantized latent $\hat{\bmy}$:
\begin{equation}
\hat{\bmx}=g_s(\hat{\bmy};\bmtheta_{g_s}). 
\end{equation}

Finally, we optimize the following training objectives:
\begin{equation}
\argmin{\bmtheta_{g_a}, \bmtheta_{h_a}, \bmtheta_{g_s}, \bmtheta_{h_s}, \{\bmtheta_{e_i}\}_{i=0}^{S-1}} \lambda \shuk{\bmx-\hat{\bmx}}^2 + R(\hat{\bmz}) + R(\hat{\bmy}),
\end{equation}
where $\lambda$ is the Lagrangian multiplier to control the rate-distortion trade-off. 

\subsubsection{Visual State Space (VSS) Block} \label{subsubsec:VSSBlock}
Inspired by \citet{liu2024vmamba}, for each of the nonlinear transforms $g_a$, $g_s$, $h_a$ and $h_s$, we design a Visual State Space (VSS) block following each upsampling or downsampling operation in the middle of the transform. 
\Cref{VSSB} illustrates the structure. 
To be specific, each VSS Block is composed of multiple VSS layers. 
Following Mamba \cite{gu2023mamba}, the VSS layer adopts a gated structure with two branches after layer normalization (LN) \cite{ba2016layer}. 
Given an input feature map $\bmf_\mathrm{in}\in\bbR^{H \times W \times C}$, the main branch processes it by:
\begin{equation}
\bmf_\mathrm{hidden}=\mathrm{LN}_2(\mathrm{2DSS}(\sigma(\mathrm{DWConv}(\mathrm{Linear}_1(\mathrm{LN}_1(\bmf_\mathrm{in})))))), 
\end{equation}
where $\mathrm{LN}$ denotes layer normalization. 
$\mathrm{2DSS}$ denotes the 2D selective scan module, which will be elaborated in \Cref{subsubsec:2DSS}.
$\sigma$ denotes the SiLU activation \cite{ramachandran2017searching}.
$\mathrm{DWConv}$ denotes the depthwise convolution. 
$\mathrm{Linear}$ denotes learnable linear projection. 

Analogously, the gating branch computes the weight vector by:
\begin{equation}
\bmw=\sigma(\mathrm{Linear}_2(\mathrm{LN}_1(\bmf_\mathrm{in}))).
\end{equation}

Finally, the two branches are combined to produce the output feature map:
\begin{equation}
\bmf_\mathrm{out}=\mathrm{Linear}_3(\bmf_\mathrm{hidden}\odot\bmw) + \bmf_\mathrm{in}, 
\end{equation}
where $\odot$ denotes the element-wise product. 

\subsubsection{2D Selective Scan (2DSS)} \label{subsubsec:2DSS}
Vanilla Mamba~\cite{gu2023mamba} can only process 1D sequences, which can not be directly applied to 2D image data. 
To effectively model spatial context, we expand 4 unfolding for selective scanning. 
Concretely, for the feature map $\bmf\in\bbR^{H\times W\times C}$, where $\bmf[h][w]\in\bbR^C$ denotes the token in the $h$-th ($0\le h<H$) row and $w$-th ($0\le w<W$) column of the feature map, the unfolding patterns are defined by
\begin{align}
\bms_1[i] &= \bmf [i \bmod W ][\lfloor i/W \rfloor], \\
\bms_2[i] &= \bmf [(N - i - 1) \bmod W][\lfloor (N - i - 1)/W \rfloor], \\
\bms_3[i] &= \bmf [\lfloor i/H \rfloor][i \bmod H], \\
\bms_4[i] &= \bmf [\lfloor (N - i - 1)/H \rfloor][(N - i - 1) \bmod H],
\end{align}
where $N=H\times W$, $0\le i<N$. $\bms_1,\bms_2,\bms_3,\bms_4\in\bbR^{N\times C}$ are the expanded and flattened token sequences.
For each flattened token sequence, we apply an S6 \cite{gu2023mamba} operator for selective scanning, producing contextual token sequences $\bms_1',\bms_2',\bms_3',\bms_4'\in\bbR^{N\times C}$. 

We then apply reversed operations to the contextual token sequences by the following folding patterns:
\begin{align}
\bmf_1'[i][j] &= \bms'_1[j \times W + i] , \\
\bmf_2'[i][j] &= \bms'_2[N - 1 - j \times W - i], \\
\bmf_3'[i][j] &= \bms'_3[i \times H + j] , \\
\bmf_4'[i][j] &= \bms'_4[N - 1 - i \times H - j],
\end{align}
where $\bmf'_1,\bmf'_2,\bmf'_3,\bmf'_4\in\bbR^{H\times W\times C}$ denote the expanded and transformed feature map of $\bmf$. 

In the end, we merge the transformed feature maps to obtain the output feature map:
\begin{equation}
    \bmf'= \bmf'_1 + \bmf'_2 + \bmf'_3 + \bmf'_4.
\end{equation}

\subsubsection{Extension to Video Compression} 
We also extend \modelname{} to video compression to explore its potential. 
Here we choose the scale-space flow (SSF) \cite{agustsson2020scale}, a renowned learned video compression model, as the base framework for extension. 
We upgrade the CNN-based transforms in 3 parts (\ie I-frame compression, scale-space flow, and residual) of 
SSF with the developed VSS blocks. 
We call this extension by \modelname{}-SSF. 
We will show and discuss the experimental results in \Cref{subsec:vid_compress_exp}.
\vspace{-0.3cm}
\section{Experiments}
\label{sec:experiments}
\subsection{Experimental Setup}
\subsubsection{Datasets and Training Details} \label{subsubsec:dataset_training_details}
For image compression, we select the Flickr30k dataset in \cite{young2014image}, consisting of 31,783 images. Each model is trained for 2M steps. For the first 1.2M steps, each batch consists of 8 randomly cropped 256$\times$256 images; for the next 0.8M steps, each batch includes 2 randomly selected 512$\times$512 upsampled images. The learning rate starts at 10\textsuperscript{-4} and drops to 10\textsuperscript{-5} at 1.8M steps, finally drops to 10\textsuperscript{-6} at 1.95M steps. We employe $\lambda \in \{0.0035,0.0067,0.013,0.025,0.05\}$ in rate-distortion loss.

For video compression, 
models are all trained on Vimeo-90k~\cite{xue2019video} for 1M steps at a learning rate of 10\textsuperscript{-4} and an additional 0.6M steps at 10\textsuperscript{-5}. In the first phase, each batch contains 8 randomly cropped 256$\times$256 images; in the second phase, each batch contains 8 randomly cropped 384$\times$256 images. We optimize video model for MSE distortion metric. In particular, we use $\lambda \in \{0.00125,0.0025,0.005,0.01,0.02, 0.04, 0.08, 0.16, 0.32\}$. Inspired by \cite{jaegle2021perceiver,meister2018unflow}, we process each video sequence in original and reversed order respectively during each optimization step.
\begin{figure}[t]
	\centering
	\subfigure
    {\includegraphics[width=.4965\textwidth]{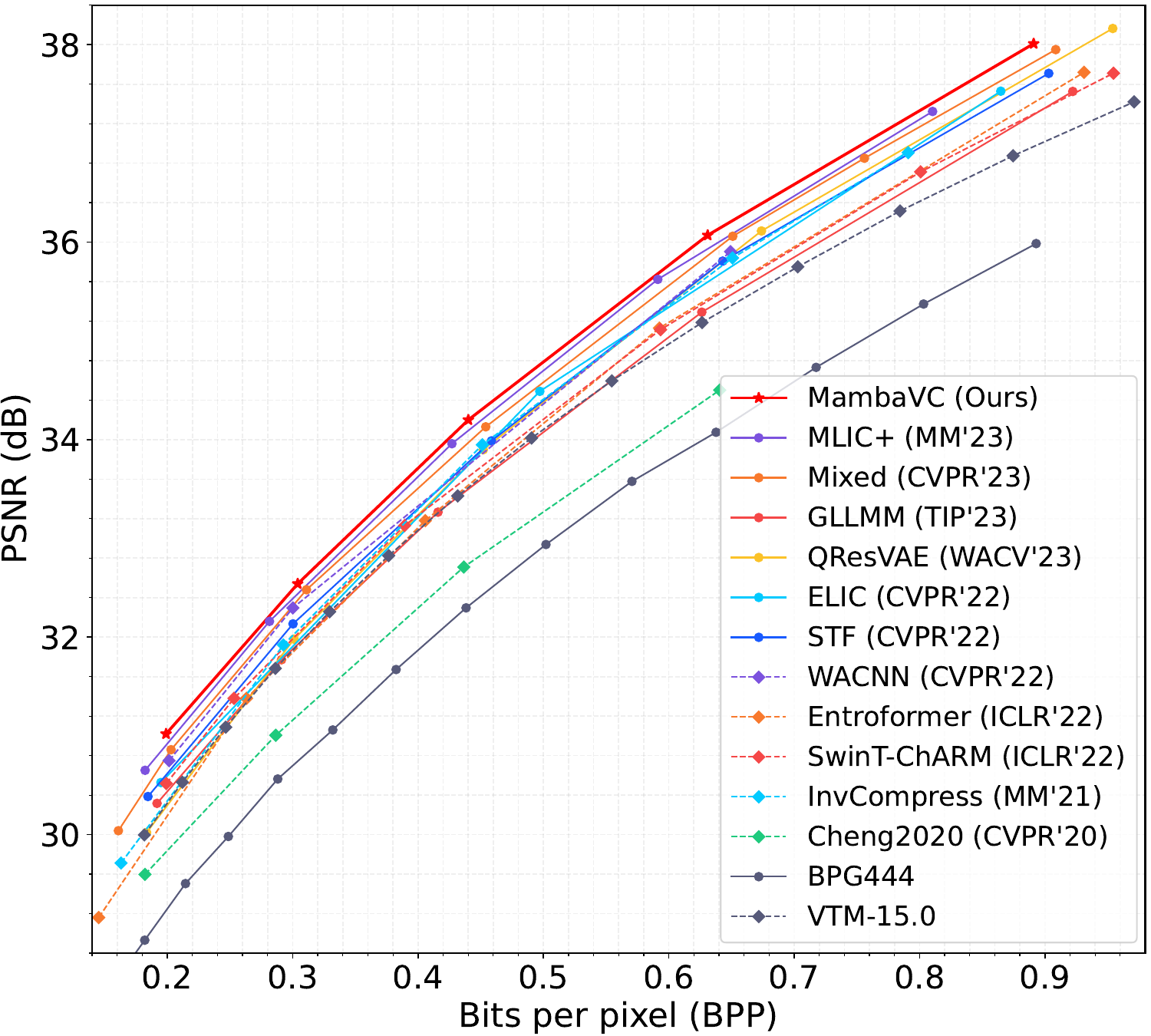}}
	\subfigure    
    {\includegraphics[width=.4965\textwidth]{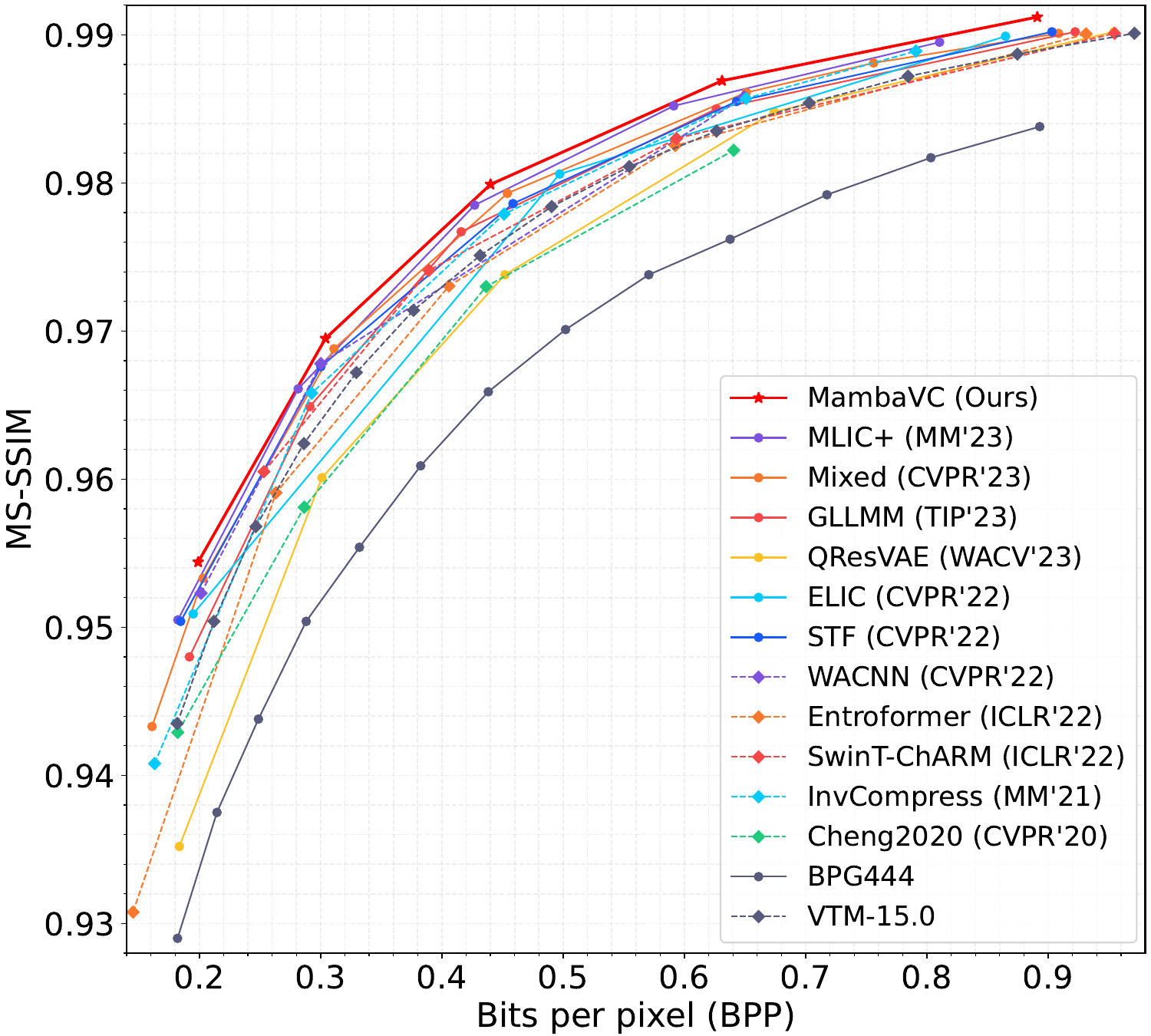}}
    \vspace{-0.2cm}
	\caption{Performance evaluation (optimized by MSE) on the Kodak dataset~\cite{franzen1999kodak}.}
	\label{main_exp}
    \vspace{-0.45cm}
\end{figure}
\subsubsection{Baselines} We conduct a comprehensive and thorough evaluation of \modelname{} in two directions on Kodak~\cite{franzen1999kodak}, CLIC2020~\cite{toderici2020clic}, JPEG-AI~\cite{jpegai} and UHD~\cite{zhang2021benchmarking} with different image resolution. First, we compare it with state-of-the-art methods, including MLIC+~\cite{jiang2023mlic}, Mixed~\cite{liu2023learned}, GLLMM~\cite{fu2023learned}, QResVAE~\cite{duan2023lossy}, ELIC~\cite{he2022elic}, STF~\cite{zou2022devil}, WACNN~\cite{zou2022devil}, Entroformer~\cite{qian2021entroformer}, Swin-ChARM~\cite{zhu2021transformer}, Invcompress~\cite{xie2021enhanced} and traditional coding method BPG444~\cite{bellard2018bpg} and VTM-15.0~\cite{bross2021overview}. Secondly, we validate the superiority of \modelname{} over its convolutional and Transformer variants in terms of performance and efficiency. Specifically, we replace the VSS Block in \modelname{} with swin transformer and GDN layer, respectively, naming them SwinVC and ConvVC. Detailed structures are shown in \Cref{Model Configurations}.

Meanwhile, we evaluate variant SSF on MCL-JCV~\cite{wang2016mcl} and UVG~\cite{mercat2020uvg}, comparing it with standard codecs AVC(x264), HEVC(x265) and the test model implementation of HEVC, called HEVC (HM). All methods fix the GOP size to 12.
\vspace{-0.2cm}
\subsection{Standard Image Compression}
\subsubsection{Comparison with the State-of-the-art Methods}
\vspace{-0.1cm}
The rate-distortion performance on Kodak dataset~\cite{franzen1999kodak} is shown in Figure~\ref{main_exp}. For fairness, all shown learned methods are optimized for minimizing MSE. Both PSNR and MS-SSIM are tested to demonstrate the robustness of \modelname{}. Compared to the previous best methods MLIC+~\cite{jiang2023mlic}, our approach yields an average PSNR improvement of 0.1 dB, while requiring only half the computational complexity and 60\% of the memory overhead as shown in Figure~\ref{fig:BDrate_MAC_size}.
\vspace{-0.2cm}
\subsubsection{Comparison of Variants}
\vspace{-0.2cm}
\label{Variant Visual Compression Performance}
The RD curves for all variants on Kodak~\cite{franzen1999kodak} are shown in \Cref{Variant_RD_curve}. To provide a clearer comparison of the performance among different variants, Figure~\ref{Rate saving} illustrates the percentage of rate savings relative to VTM-15.0 for achieving equivalent PSNR.
Figure~\ref{Var_performance} demonstrates that \modelname{} consistently outperforms SwinVC and ConvVC in various scenarios. SwinVC, as highlighted in previous work, surpasses ConvVC. Both \modelname{} and SwinVC exhibit higher compression efficiency compared to VTM-15.0, whereas ConvVC falls short. As the rate increase, SwinVC's performance advantage slightly diminishes, while \modelname{} remains unaffected. In Table~\ref{BDrate over VTM}, we present the BD-rate of different variants compared to VTM-15.0 across four datasets. \modelname{} achieves an average bitrate savings of 13.35\%, while SwinVC achieves an average savings of 1.94\%. In contrast, ConvVC consumes an average of 4.76\% more bits. Notably, \modelname{} is the only variant that surpasses VTM-15.0 on UHD~\cite{zhang2021benchmarking}, highlighting its potential for high-resolution images, which will be discussed in the next section.
See \Cref{Variant Visual Compression Performance on Different Datasets} for further details. 
\begin{table}[h!]
    \centering
    \caption{BD-rate (lower is better) of the variants, with VTM-15.0 as the anchor.}
    \begin{tabular}{lcccc}
        \toprule
        Method & Kodak~\cite{franzen1999kodak} & CLIC2020~\cite{toderici2020clic} & JPEG-AI~\cite{jpegai} & UHD~\cite{zhang2021benchmarking} \\ 
        \midrule
        BPG444 & 29.85\% & 32.99\% & 43.87\% & 20.87\% \\ 
        \midrule
        ConvVC & 1.59\% & 0.13\% & 4.02\% & 11.50\% \\ 
        SwinVC & -6.08\% & -5.69\% & -0.61\% & 8.59\% \\
        \modelname{} & \textbf{-15.41\%} & \textbf{-16.68\%} & \textbf{-12.36\%} & \textbf{-5.95\%} \\    
        \bottomrule
    \end{tabular}
    \label{BDrate over VTM}
\end{table}

\vspace{-0.2cm}

\subsection{High-Resolution Image Compression}
\vspace{-0.15cm}
\begin{wraptable}{r}{0.5\textwidth}
    \vspace{-0.68cm}
  \centering
  \caption{BD-rate of \modelname{} over variants.}
  \vspace{0.15cm}
  \begin{tabular}{lcc}
    \toprule
    Datasets & SwinVC & ConvVC \\
    \midrule
    Kodak~\cite{franzen1999kodak} & -9.33\% & -15.67\% \\
    CLIC2020~\cite{toderici2020clic} & -13.65\% & -16.02\% \\
    JPEG-AI~\cite{jpegai} & -11.32\% & -16.08\% \\
    UHD~\cite{zhang2021benchmarking} & -17.18\% & -20.05\% \\
    \bottomrule
    \label{SwinVC ConvVC BD-rate}
  \end{tabular}
  \vspace{-0.23cm}
  \caption{Complexity (MACs) of different models.}
  \vspace{0.15cm}
  \setlength\tabcolsep{2pt}
  \begin{tabular}{lccc}
    \toprule
    Datasets & \modelname{} & SwinVC & ConvVC \\
    \midrule
    Kodak~\cite{franzen1999kodak} & 0.32T & 0.56T & 0.42T \\
    CLIC2020~\cite{toderici2020clic} & 7.24T & 12.45T & 9.43T \\
    JPEG-AI~\cite{jpegai} & 7.84T & 13.51T & 12.52T \\
    UHD~\cite{zhang2021benchmarking} & 18.02T & 30.98T & 23.48T \\
    \bottomrule
    \label{SwinVC ConvVC Computational complexity}
  \end{tabular}
  \vspace{-0.5cm}
\end{wraptable}
Recent work~\cite{wang2024graph,yang2024mambamil} has demonstrated Mamba's advantages in long-range modeling. To explore this potential in visual compression, we compare our \modelname{} against SwinVC and ConvVC on images of varying resolutions in two ways. Specifically, we downsample high-resolution images from the UHD~\cite{zhang2021benchmarking} by different factors to create multiple sets of images with the same distribution but different sizes. As shown in Figure~\ref{fig:image_resolution}, \modelname{} saves more bits as the resolution increases compared to the other variants. To mitigate the impact of specific dataset distributions, we test across four datasets with different resolutions. As indicated in Table~\ref{SwinVC ConvVC BD-rate}, the performance advantage of \modelname{} on the high-resolution UHD~\cite{zhang2021benchmarking} is significantly greater than on the lower-resolution Kodak~\cite{franzen1999kodak}. For datasets with similar sizes, like CLIC2020~\cite{toderici2020clic} and JPEG-AI~\cite{jpegai}, the performance advantage is relatively consistent. We also record the change in computational cost across different resolutions. As shown in Table~\ref{SwinVC ConvVC Computational complexity}, with increasing image sizes, the computational gap widened from an initial 0.23 TMACs and 0.1 TMACs to a final 12.96 TMACs and 5.46 TMACs, separately. These results indicate that \modelname{} has a distinct advantage in compressing high-resolution images. This potential may influence the future development of specialized fields such as medical imaging and satellite imagery.
\begin{figure}[h]
    \vspace{-0.3cm}
	\centering
	\subfigure[RD curve on MCL-JCV~\cite{wang2016mcl}.\label{MCL-JCV}]
    {\includegraphics[width=.4965\textwidth]{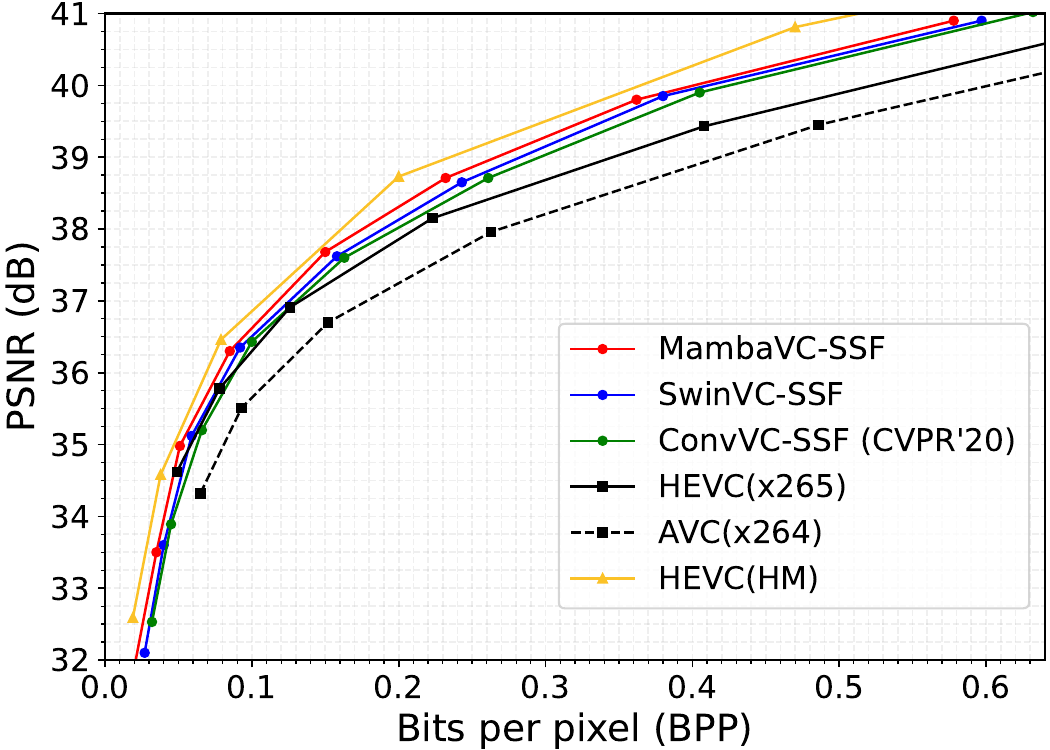}}
	\subfigure[RD curve on UVG~\cite{mercat2020uvg}.\label{UVG}] 
    {\includegraphics[width=.4965\textwidth]{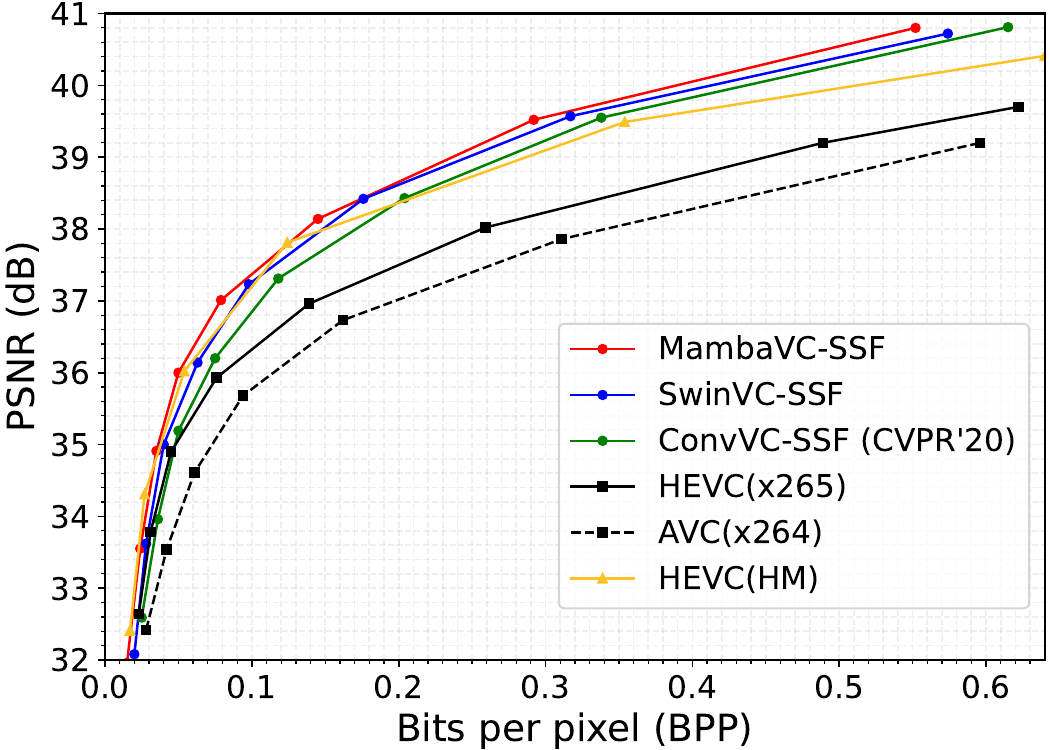}}
	\caption{Video compression performance evaluation on benchmark datasets.}
	\label{Video RD curve}
    \vspace{-0.5cm}
\end{figure}

\subsection{Video Compression with SSF Backbone} \label{subsec:vid_compress_exp}
Following the configuration of \citet{agustsson2020scale}, we evaluated our method on the MCL-JCV~\cite{wang2016mcl} and UVG~\cite{mercat2020uvg} datasets. 
To ensure a more comprehensive comparison, we also construct the CNN- and Swin-Transformer-based counterparts with \modelname{}-SSF, denoted as SwinVC-SSF and ConvVC-SSF, respectively. 
Detailed configurations for different models can be found in \Cref{subsubsec:dataset_training_details,sec:appendix}.
Figure~\ref{Video RD curve} presents the RD curves of MambaVC-SSF with its different variants and traditional methods. The mamba-based model outperforms its convolutional and transformer counterparts. However, the performance improvement in video compression is not as pronounced as in image compression, possibly because merely changing the nonlinear transformation structure is insufficient to capture more redundancy. Additionally, all variants still fall short of HM in performance on the MCL-JCV dataset, indicating significant room for further improvement.

\subsection{Computational and Memory Efficiencies}
\vspace{-0.1cm}
\label{Computational complexity}
To explore the advantage of Mamba's linear complexity in visual compression, we evaluate the memory overhead and computational complexity on the Kodak dateset~\cite{franzen1999kodak}. As results shown in Table~\ref{all Computational complexity}, \modelname{} exhibits the best performance across different variants. While MLIC+~\cite{jiang2023mlic} incurs greater computational cost due to its adoption of a more advanced entropy model, it doesn't achieve superior performance. On the other hand, method~\cite{liu2023learned} combining convolution and transformer, while falling short in both computational and storage aspects compared to SwinVC and ConvVC, further underscores the significance of \modelname{} as a novel framework.

\setlength{\textfloatsep}{6pt}
\begin{table}[h!]
    \caption{Computational and memory efficiencies of different components. All models are trained with $\lambda=0.05$. The complexity of the entropy model is attributed to the hyper decoder $h_s$. Except for \cite{liu2023learned}, the other approaches have symmetric $g_a$ and $g_s$, so we do not repeat their presentation.}
    \centering
    \resizebox{\columnwidth}{!}{
        \setlength{\tabcolsep}{.5em}{
            \begin{tabular}{ccccccccccc}
                \toprule
                \multirow{2}{*}{Method} & \multicolumn{4}{c}{MACs} & \multicolumn{4}{c}{FLOPs} & \multirow{2}{*}{Peak memory} & \multirow{2}{*}{Model params}\\
                \cmidrule(lr){2-5} \cmidrule(lr){6-9} 
                & $g_a$ & $h_a$ & $h_s$ & total & $g_a$ & $h_a$ & $h_s$ & total & \\
                \midrule
                \modelname{} & 140.9G & 631.1M & 43.6G  & 326.1G & 362.3G & 1.4G & 89.0G  & 815.1G & 611.5M & 53.3M \\
                SwinVC  & 257.9G & 929.5M & 44.2G  & 560.9G & 517.1G & 1.8G & 93.9G  & 1.1T   & 706.6M & 60.4M \\
                ConvVC  & 188.8G & 1.6G   & 45.8G  & 425.1G & 377.8G & 3.3G & 92.6G  & 851.5G & 769.6M & 74.0M \\
                MLIC+~\cite{jiang2023mlic}   & 145.9G & 1.65G  & 210.2G & 503.6G & 292.1G & 3.2G & 422.5G & 1.0T   & 1.3G   & 116.7M \\
                Mixed~\cite{liu2023learned}   & 267.2G & 1.0G   & 46.8G  & 717.1G & 544.1G & 2.2G & 90.3G  & 1.5T   & 877.8M & 76.6M \\
                \bottomrule
            \end{tabular}
        }
    }
    \label{all Computational complexity}
\end{table}

\vspace{-0.4cm}
\subsection{Analysis}
\vspace{-0.1cm}
\label{sec:analysis}
\subsubsection{Latent Correlation and Distribution}

\begin{figure}[h]
\vspace{-0.1cm}
\centering
\includegraphics[width=0.9\textwidth]{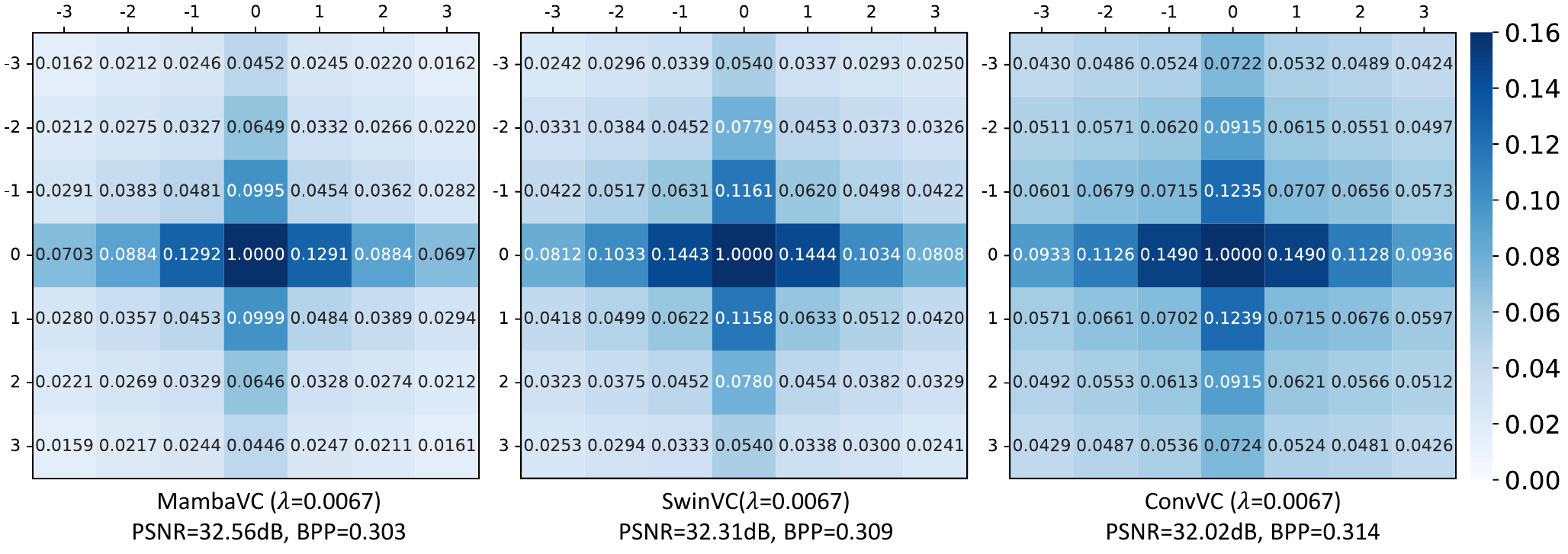}
\vspace{-0.25cm}
\caption{Latent correlation of $(\bm{y} - \bm{\mu}) / \bm{\sigma}$. All models are trained with $\lambda=0.0067$. The value at position $(i,j)$ represents cross-correlation between spatial locations $(x,y)$ and $(x+i,y+j)$ along the channel dimension, averaged across all images on Kodak~\cite{franzen1999kodak}.}  
\label{Latent correlation}
\vspace{-0.15cm}
\end{figure}

Learned visual compression redundancy removal involves two key steps: nonlinear encoding transform and using a conditionally factorized Gaussian prior distribution to decorrelate the latent $\bm{y}$. 

\begin{wrapfigure}{r}{0.45\textwidth}
    \vspace{-0.42cm}
    \centering
    \includegraphics[width=0.45\textwidth]{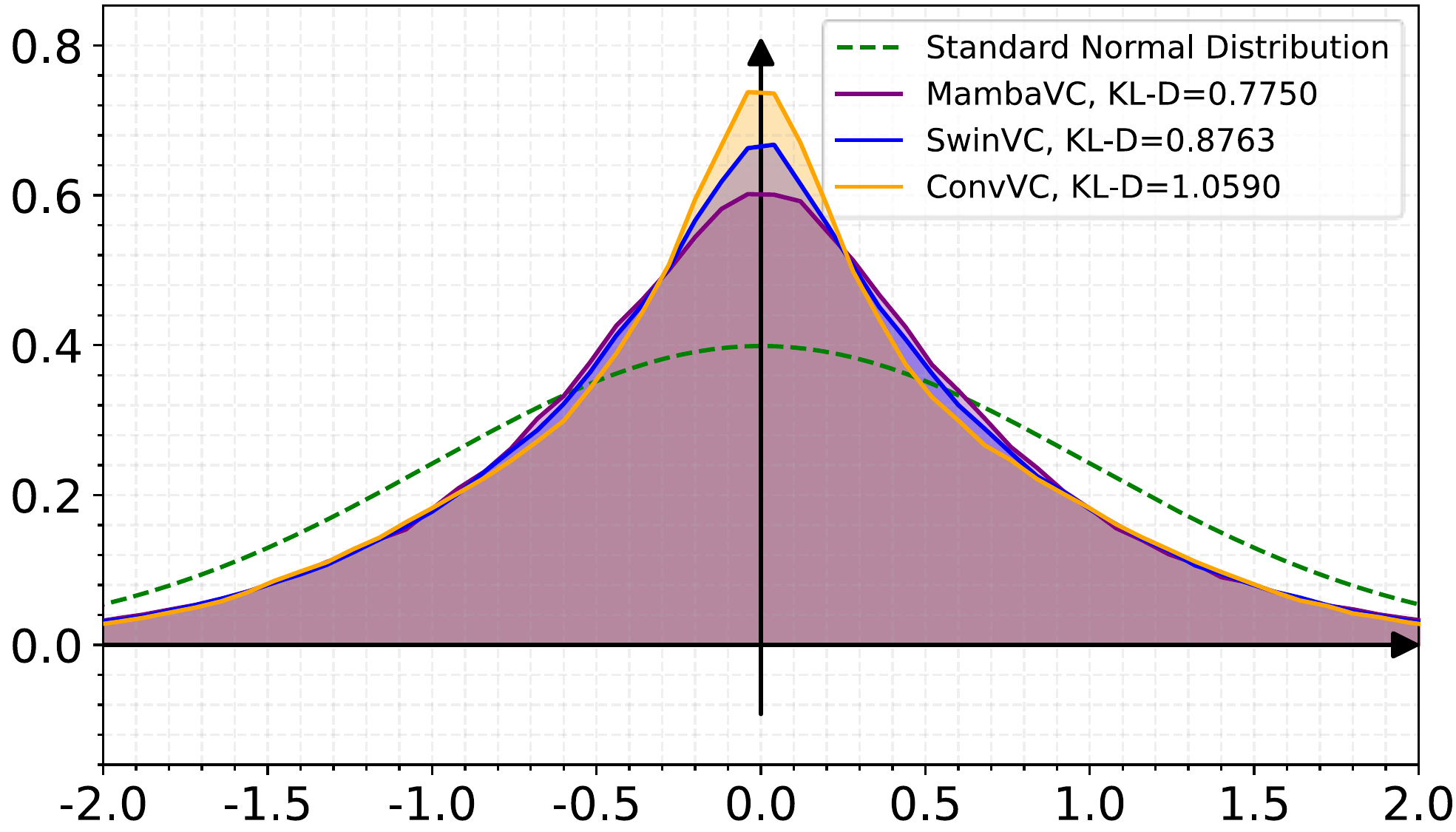}
    \vspace{-0.42cm}
    \caption{Distribution of $\bm{\ddot{y}}$. KL-D represents Kullback-Leibler Divergence~\cite{kullback1951information} compared to the standard normal distribution.}
    \label{density}
    \vspace{-0.35cm}
\end{wrapfigure}
Specifically, the former converts the input signal from the image domain to the feature domain, while the latter uses a hyper network to learn the mean and variance $(\bm{\mu}, \bm{\sigma})$ of latent $\bm{y}$, assuming a Gaussian distribution, to further reduce correlation. As various correlations and redundancies are eliminated, less information needs to be entropy coded, thereby improving compression efficiency. 
To this end, we visualized the correlation between each spatial pixel in $\bm{\ddot{y}} \triangleq (\bm{y}-\bm{\mu}) / \bm{\sigma}$ and its surrounding positions, which we refer to as latent correlation. \Cref{Latent correlation} indicates that \modelname{} has lower correlations at all distances compared to SwinVC and ConvVC. Theoretically, decorrelated latent should follow a standard normal distribution (SND). To verify this, we fit the distribution curves for different methods and calculated the KL divergence~\cite{kullback1951information} from SND, as shown in \Cref{density}. The curve for \modelname{} is noticeably closer to the SND with a smaller KL divergence~\cite{kullback1951information}, which indicates the Mamba-based hyper network can learn $(\bm{\mu}, \bm{\sigma})$ more accurately. We also investigate the hyper latent correlation and the relationship between $\lambda$ and correlation, as shown in \Cref{Hyper Latent Correlation}.
\clearpage
\begin{wrapfigure}{r}{0.68\textwidth}
    \vspace{-0.4cm}
    \centering
    \subfigure[\modelname{}] 
    {\includegraphics[width=.2\textwidth]{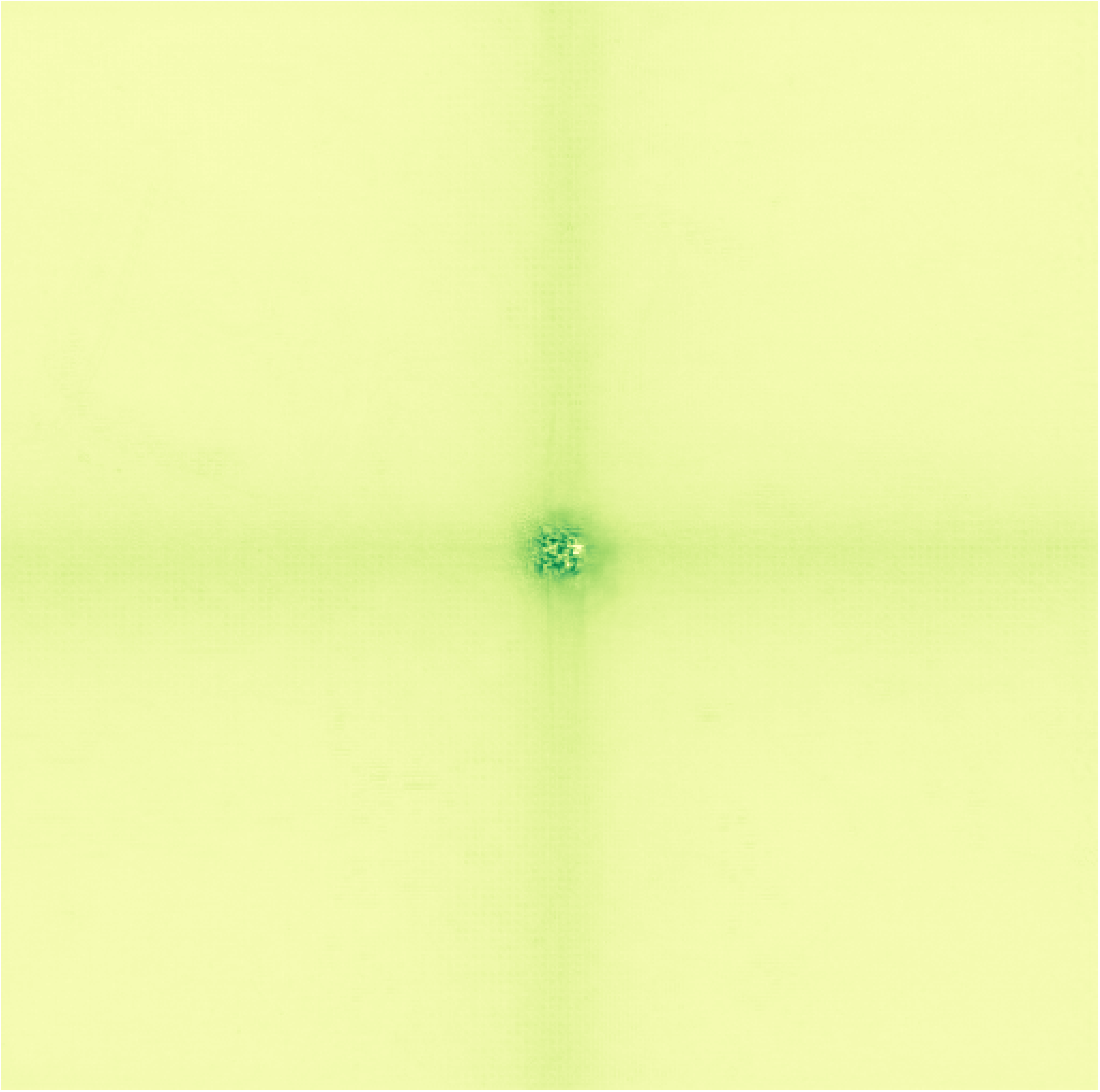}}
	\subfigure[SwinVC]     
    {\includegraphics[width=.2\textwidth]{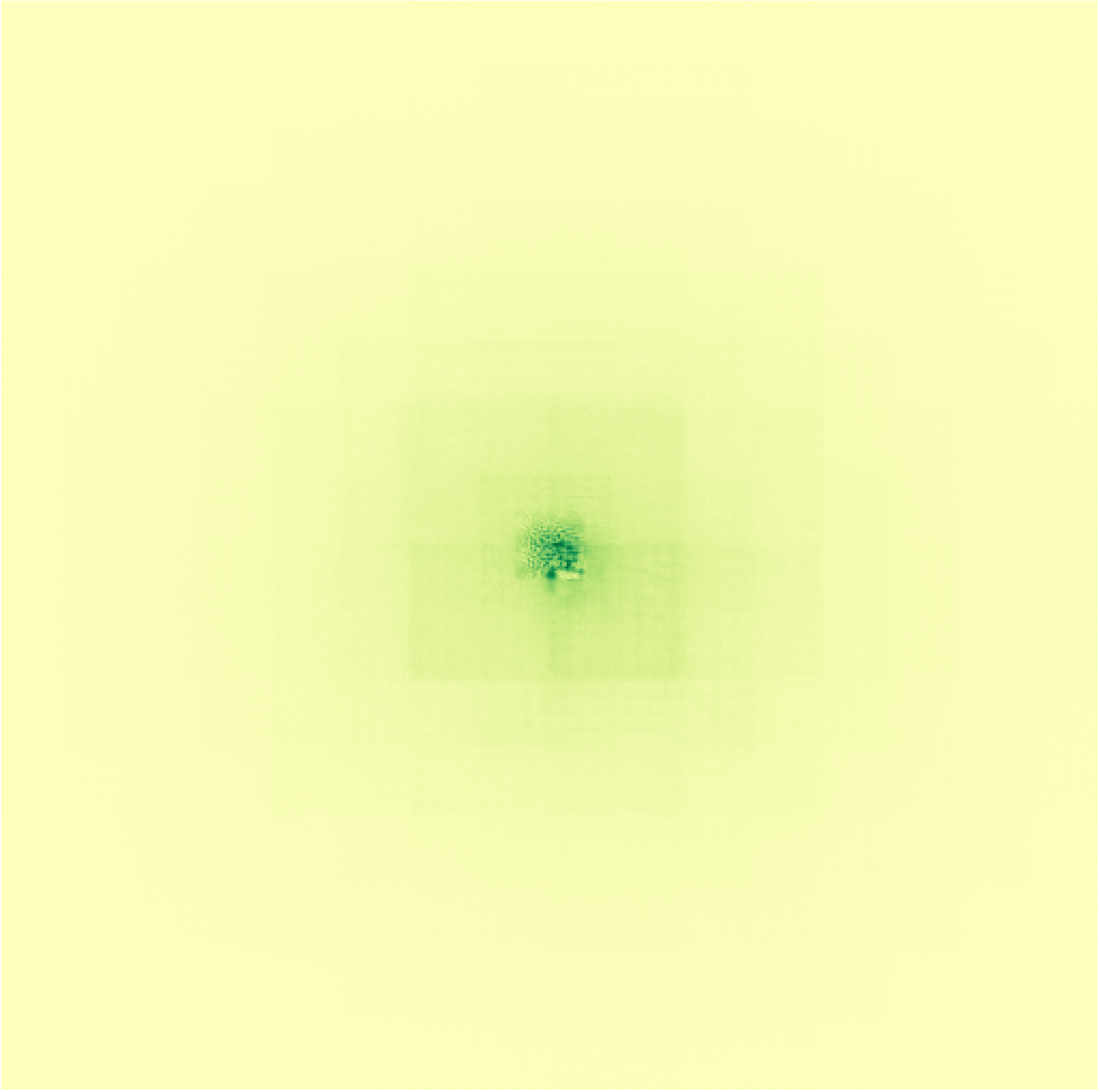}}
    \subfigure[ConvVC]     
    {\includegraphics[width=.2\textwidth]{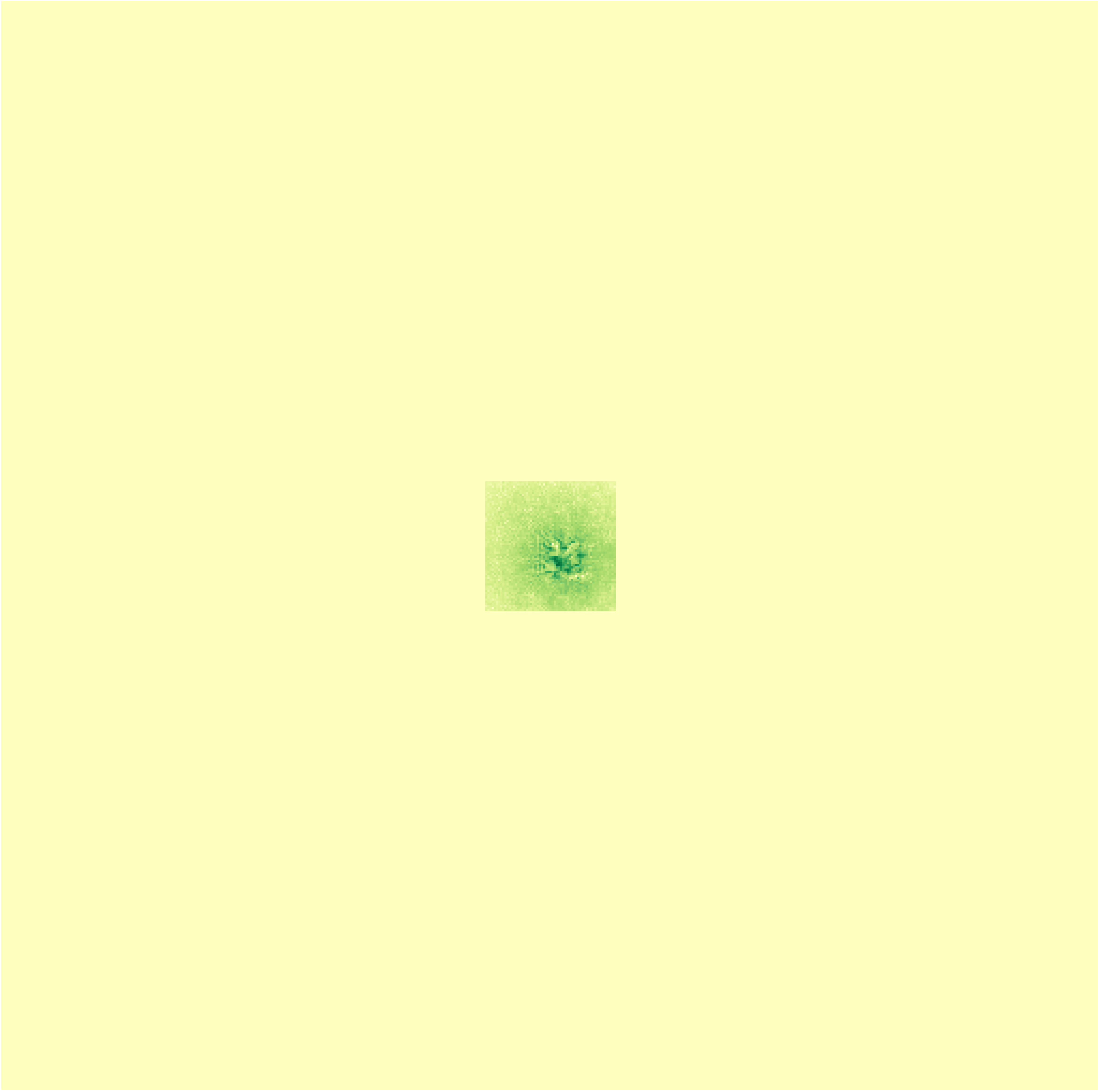}}
    \subfigure 
    {\includegraphics[width=.036\textwidth]{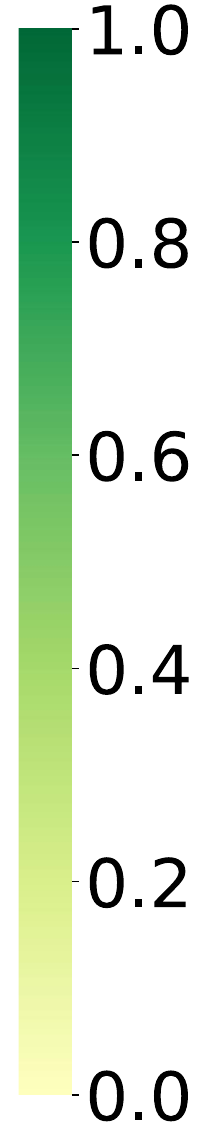}}
    \vspace{-0.1cm}
    \caption{Effective Receptive Field (ERF) of encoders $g_a$ in different models trained on Kodak~\cite{franzen1999kodak}.}
    \label{ERF}
\end{wrapfigure}


\subsubsection{Effective Receptive Field}
The effective receptive field (ERF)~\cite{luo2016understanding} denotes the region of the input that a neuron in a neural network "perceives". A larger receptive field enables the network to capture related information from a wider area. This characteristic aligns perfectly with the nonlinear encoder in visual compression, as it reduces redundancy in images through feature extraction and dimensionality reduction. Consequently, we are keenly interested in examining the receptive field sizes of \modelname{} and its variants. As shown in \Cref{ERF}, \modelname{} is the only model with a global ERF, while ConvVC has the smallest receptive field. This confirms that in high-resolution scenarios, \modelname{} can leverage more pixels globally to eliminate redundancy, whereas SwinVC and ConvVC, with their limited receptive fields, can only utilize local information, leading to performance differences.
\vspace{-0.2cm}
\subsubsection{Quantize Deviation}
\begin{figure}[h]
    \vspace{-0.25cm}
    \centering
    \includegraphics[width=\textwidth]{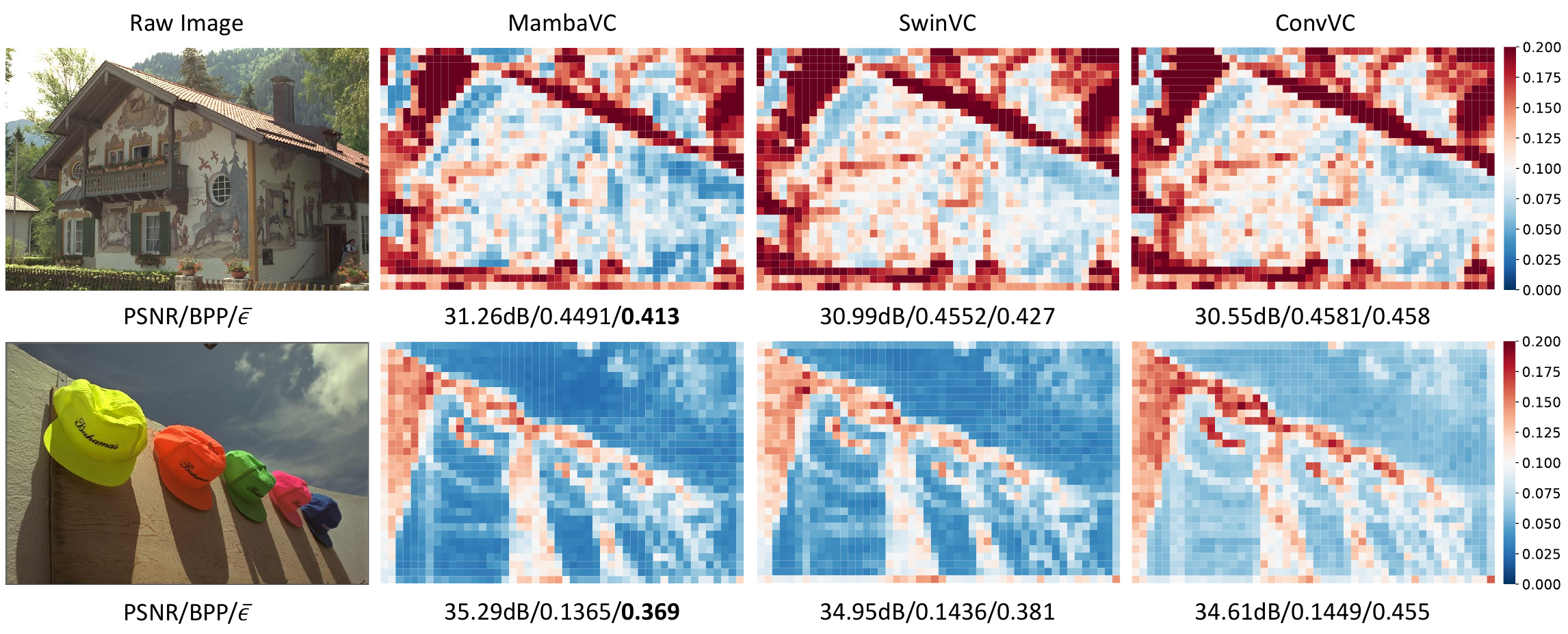}
    \vspace{-0.4cm}
    \caption{Scaled deviation map of \textit{kodim03} and \textit{kodim24} for \modelname{}, SwinVC and ConvVC.}
    \label{deviation}
    \vspace{-0.3cm}
\end{figure}

\vspace{-0.1cm}
In lossy compression, quantization is the primary source of information loss. According to \cite{xie2021enhanced}, we assess this loss by examining the deviation $\overline{\epsilon}$ between the latent $\bm{y}\in\mathbb{R}^{H \times W \times C}$ and its quantized counterpart $\bm{\hat{y}}\in\mathbb{R}^{H \times W \times C}$. \Cref{deviation} presents the scaled deviation map and specific values. Each pixel in the deviation map is the mean of the absolute deviation along the channel dimension after scaling. Compared to \modelname{}, SwinVC and ConvVC exhibit an average increase in information loss of 3.3\% and 17\%, respectively. The visualized results also indicate that \modelname{} has smaller information loss at the majority of positions (deeper blue and lighter red).

\vspace{-0.4cm}
\section{Conclusions}
\label{sec:conclusion}
\vspace{-0.15cm}
In this paper, we introduced \modelname{}, the first visual compression network based on the state-space model. 
\modelname{} built a visual state space (VSS) block with 2D selective scanning (2DSS) mechanism to improve global context modeling and content compression. 
Experimental results showed that \modelname{} achieves superior rate-distortion performance compared to CNN and Transformer variants while maintaining computational and memory efficiencies. 
These advantages are even more pronounced with high-resolution images, highlighting \modelname{}'s potential and scalability in real-world applications. 
Compared to other designs, \modelname{} exhibits stronger redundancy elimination, larger receptive fields, and lower quantization loss, revealing its comprehensive advantages for compression. 
We hope \modelname{} can offer a basis for exploring SSMs in compression and inspire future works.


\bibliographystyle{plainnat}
\bibliography{main}

\newpage
\appendix
\label{sec:appendix}
\section{Model Configurations}
\label{Model Configurations}
\subsection{Our Method}
\textbf{MambaVC} 
The detailed architecture has been delineated in Section~\ref{The proposed \modelname{}}. For the number of channels and layers, we set them as $(C_1, C_2, C_3, C_4, C_5, C_6)=(256, 256, 256, 320, 256, 192)$ and $(L_1, L_2, L_3, L_4)=(2, 2, 9, 2)$, respectively. Due to the high resolution of images in UHD, which slows down inference, we randomly select 20 images from the UHD dataset and crop their length to 3328 pixels along the center for use as the test set.

\textbf{MambaVC-SSF} 
For encoder/decoder and hyper encoder/decoder in SSF~\cite{agustsson2020scale}, there is a VSS Block following each upsampling or downsampling operation, except when generating the reconstructed image or latent with layer number $(L_1, L_2, L_3, L_4, L_5, L_6)=(1, 2, 3, 1, 1, 1)$.

\subsection{Convolutional Variant}
\begin{figure}[h]\label{fig:convvc}
\centering
\includegraphics[width=0.9\textwidth]{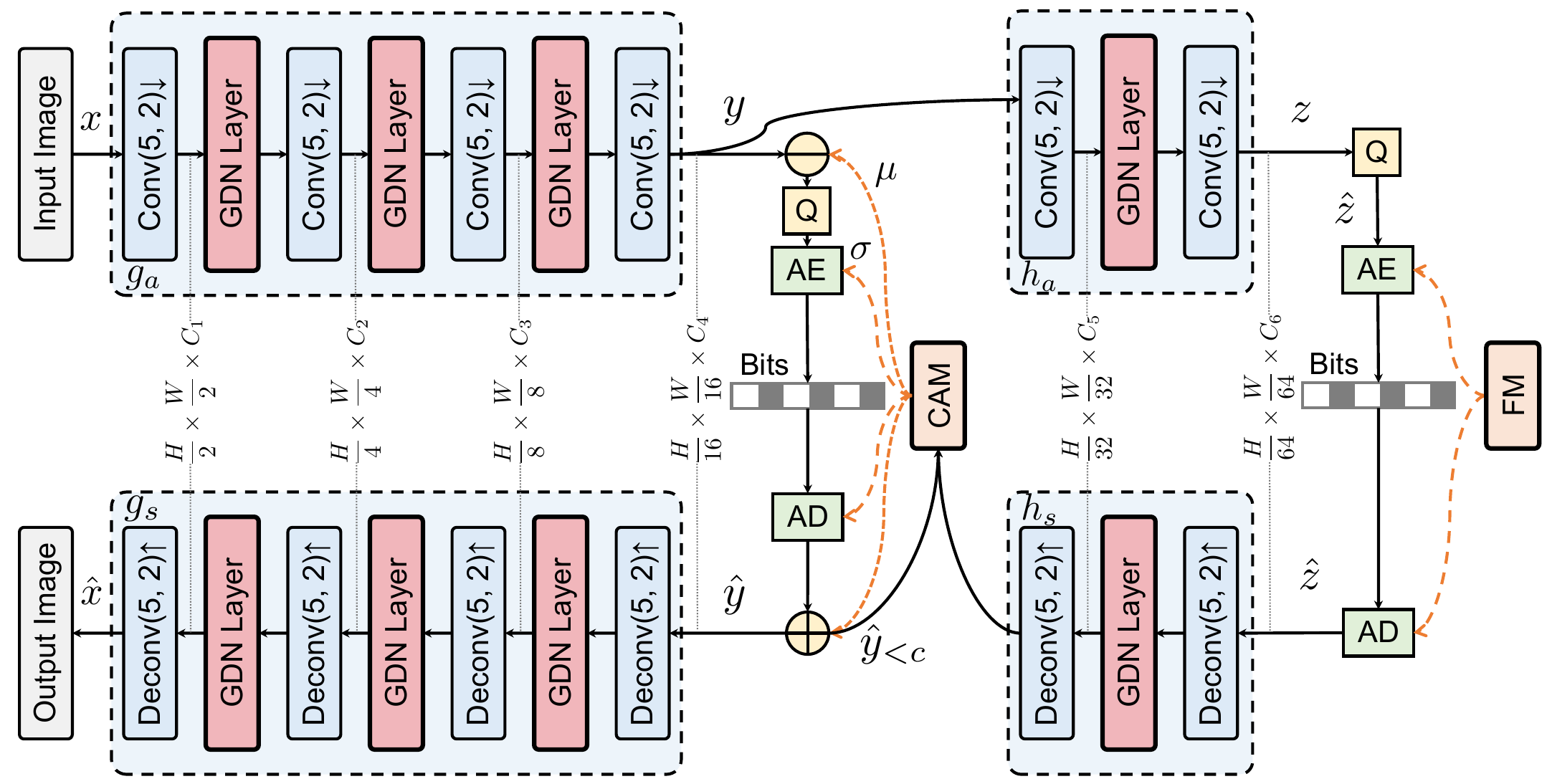}
\caption{Architecture of ConvVC.}  
\end{figure}

\textbf{ConvVC} The architecture of ConvVC are shown in Figure~\ref{fig:convvc}. Specifically, we replaced the VSS Block with the popular GDN layer~\cite{balle2016density}, which has been proven effective in Gaussianizing the local joint statistics of natural images. To compensate for the limited effective receptive field of convolutions, we set all convolutional kernels to a size of 5. For architecture, our base model has the following parameters: $(C_1, C_2, C_3, C_4, C_5, C_6)=(448, 448, 448, 320, 448, 192)$.

\subsection{Transformer Variant}
\begin{figure}[h]\label{fig:swinvc}
\centering
\includegraphics[width=0.9\textwidth]{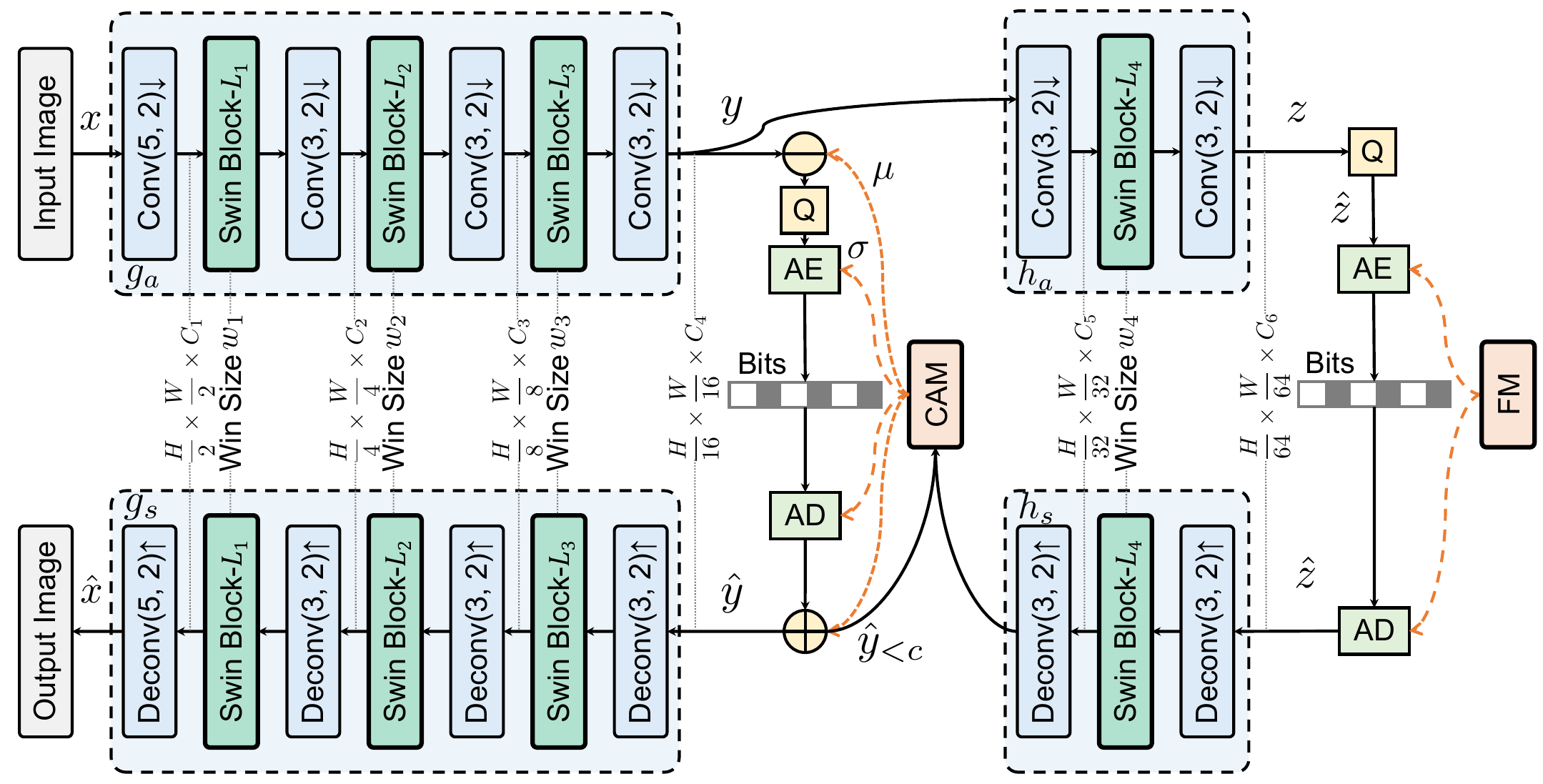}
\caption{Architecture of SwinVC.}  
\end{figure}

\textbf{SwinVC} 
Among a large number of vision transformer variants, we select Swin Transformer~\cite{dosovitskiy2020image} as network components for its lower complexity and superior modeling capability. As shown in Figure~\ref{fig:swinvc}, the layer number $(L_1, L_2, L_3, L_4)=(2, 2, 9, 2)$ and  window size $(w_1, w_2, w_3, w_4)=(8, 8, 8, 4)$ are common to all experiments. For channels, we set $(C_1, C_2, C_3, C_4, C_5, C_6)=(256, 256, 256, 320, 256, 192)$. 

\textbf{SwinVC-SSF} 
The original downsampling modules remain untouched. Following the structure akin to the image model, we utilize the Swin Transformer~\cite{dosovitskiy2020image}, albeit without any LayerNorm, instead appending a ReLU layer afterward. Both latent and hyper latent channels are set at 192. For I-frame compression, scale-space flow, and residual, we employ window sizes of 8, 4, and 8, respectively. The layer number is the same as MambaVC-SSF.

\section{Classical Standards}
In this section, we provide the evaluation script used for traditional methods.

\subsection{Image Compression}
\textbf{BPG444:} We get BPG software from {\url{http://bellard.org/bpg/}} and use command as follows:
\begin{verbatim}
bpgenc -e x265 -q [quality] -f 444 
-o [encoded bitstream file] [input image file]
bpgdec -o [output image file] [encoded bitstream file]
\end{verbatim}

\textbf{VTM-15.0:} VTM is sourced from {\url{https://vcgit.hhi.fraunhofer.de/jvet/VVCSoftware_VTM}}. The command is:
\begin{verbatim}
VVCSoftware_VTM/bin/EncoderAppStatic -i [input YUV file] -c [config file] 
-q [quality] -o /dev/null -b [encoded bitstream file] 
-wdt 1976 -hpt 1312 -fr 1 -f 1 
--InputChromaFormat=444 --InputBitDepth=8 --ConformanceWindowMode=1
VVCSoftware_VTM/bin/DecoderAppStatic -b [encoded bitstream file] 
-o [output YUV file] -d 8
\end{verbatim}

\subsection{Video Compression}
\textbf{AVC(x264)}

\begin{verbatim}
ffmpeg -y -pix_fmt yuv420p -s [resolution] -r [frame-rate] -crf [quality]
-i [input yuv420 raw video] -c:v libx264 -preset medium -tune zerolatency
-x264-params "keyint=12:min-keyint=12:verbose=1" [output mkv file path]
\end{verbatim}

\textbf{HEVC(x265)}

\begin{verbatim}
ffmpeg -pix_fmt yuv420p -s [resolution] -r [frame-rate] -tune zerolatency
-y -i [input video] -c:v libx265 -preset medium -crf [quality] 
-x265-params "keyint=12:min-keyint=12:verbose=1" [output file path]
\end{verbatim}

\textbf{HEVC(HM)}

\begin{verbatim}
HM/bin/TAppEncoderStatic -c HM/cfg/encoder_lowdelay_P_main.cfg 
-i [input video] --InputBitDepth=8 -wdt [width] 
-hgt [height] -fr [frame-rate] -f [frames number] 
-o [output video] -b [encoded bitstream file] -ip 12 -q [quality]
\end{verbatim}
\newpage
\section{More Results}

\subsection{Variant Visual Compression Performance on Different Datasets}
\label{Variant Visual Compression Performance on Different Datasets}
\begin{figure}[h]
	\centering
	\subfigure[\modelname{} Variant RD comparison.\label{Variant_RD_curve}]
    {\includegraphics[width=.4965\textwidth]{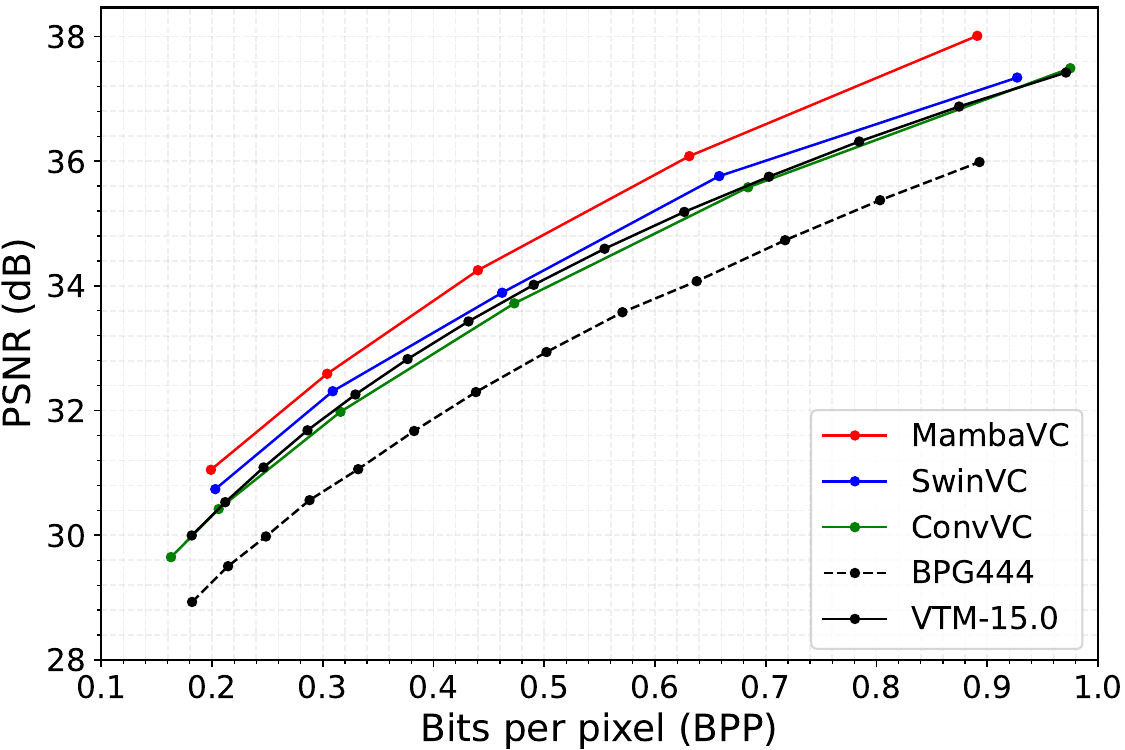}}
	\subfigure[Rate saving (\%) over VTM-15.0 (larger is better).\label{Rate saving}] 
    {\includegraphics[width=.4965\textwidth]{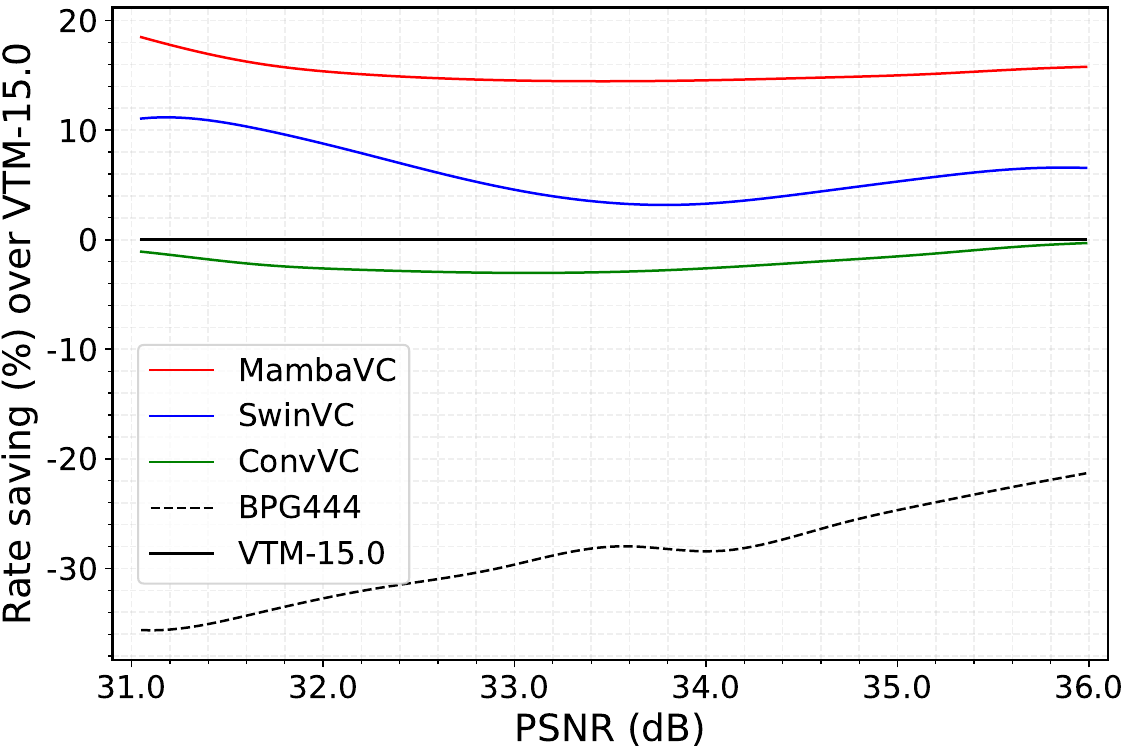}}
    \vspace{-0.3cm}
	\caption{Comparison of compression efficiency on Kodak~\cite{franzen1999kodak} among different variants.}
	\label{Var_performance}
\end{figure}
\vspace{-0.4cm}
\begin{figure}[h]
	\centering
	\subfigure[\modelname{} Variant RD comparison.]
    {\includegraphics[width=.4965\textwidth]{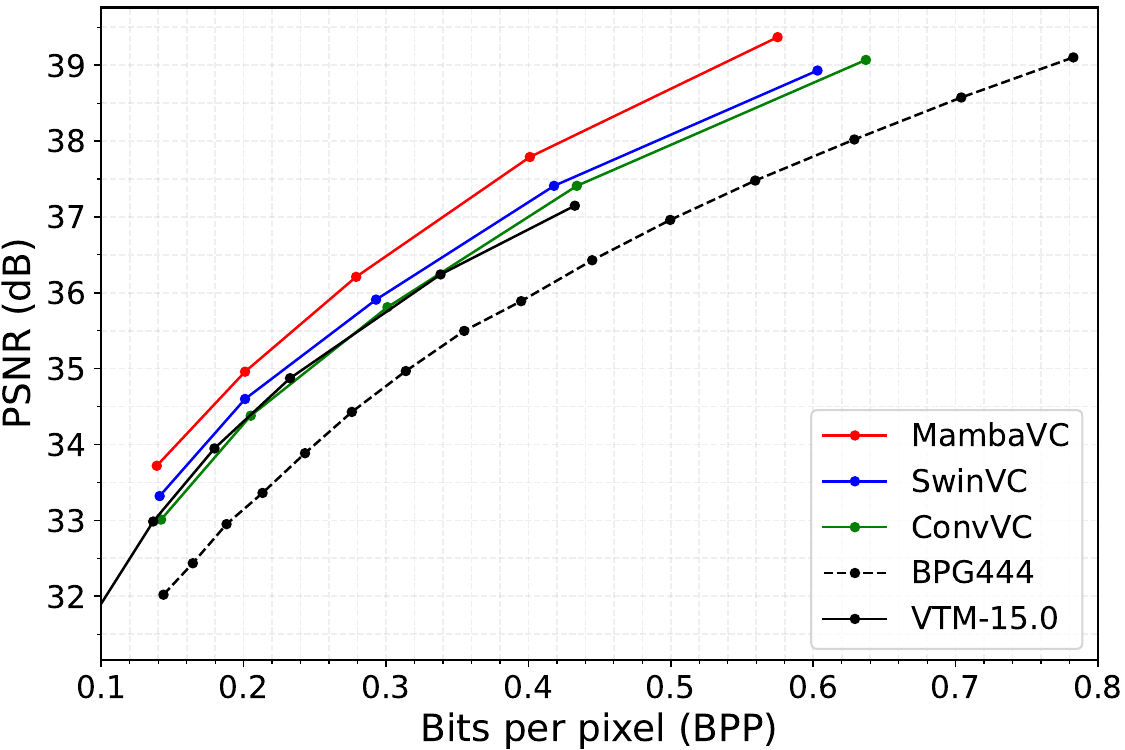}}
	\subfigure[Rate saving (\%) over VTM-15.0 (larger is better).] 
    {\includegraphics[width=.4965\textwidth]{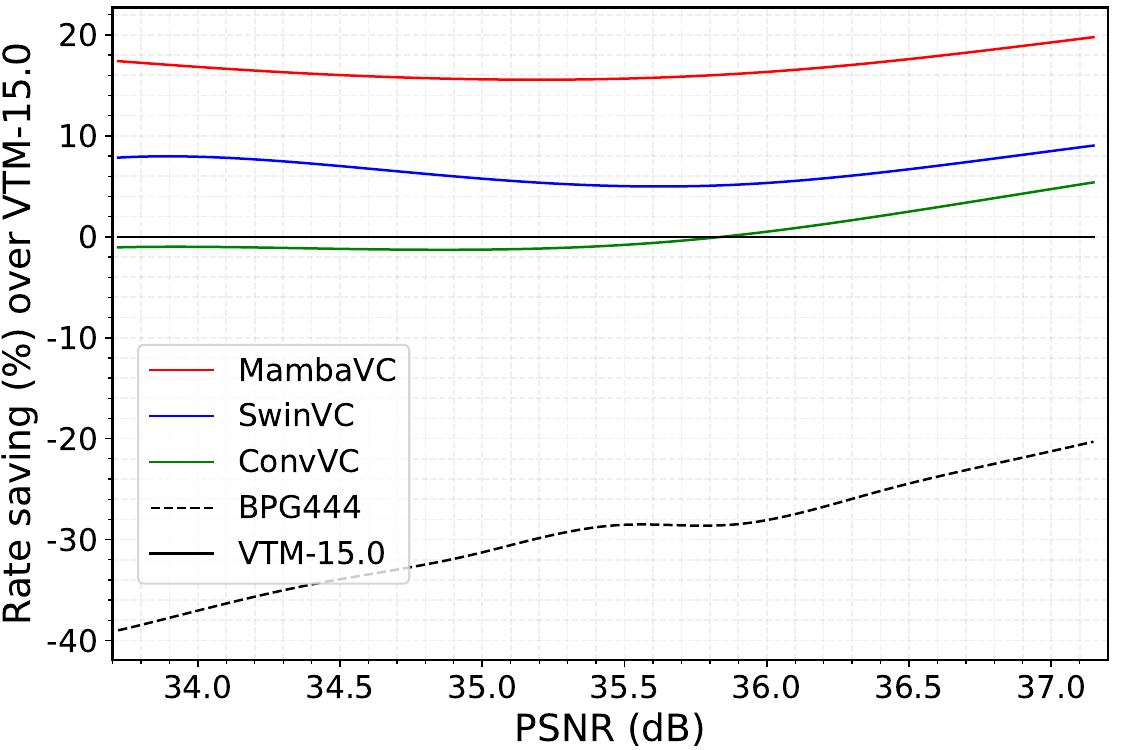}}
    \vspace{-0.3cm}
	\caption{Comparison of compression efficiency on CLIC2020~\cite{toderici2020clic} among different variants.}
	\label{CLIC_Var_performance}
\end{figure}
\vspace{-0.4cm}
\begin{figure}[h]
	\centering
	\subfigure[\modelname{} Variant RD comparison.]
    {\includegraphics[width=.4965\textwidth]{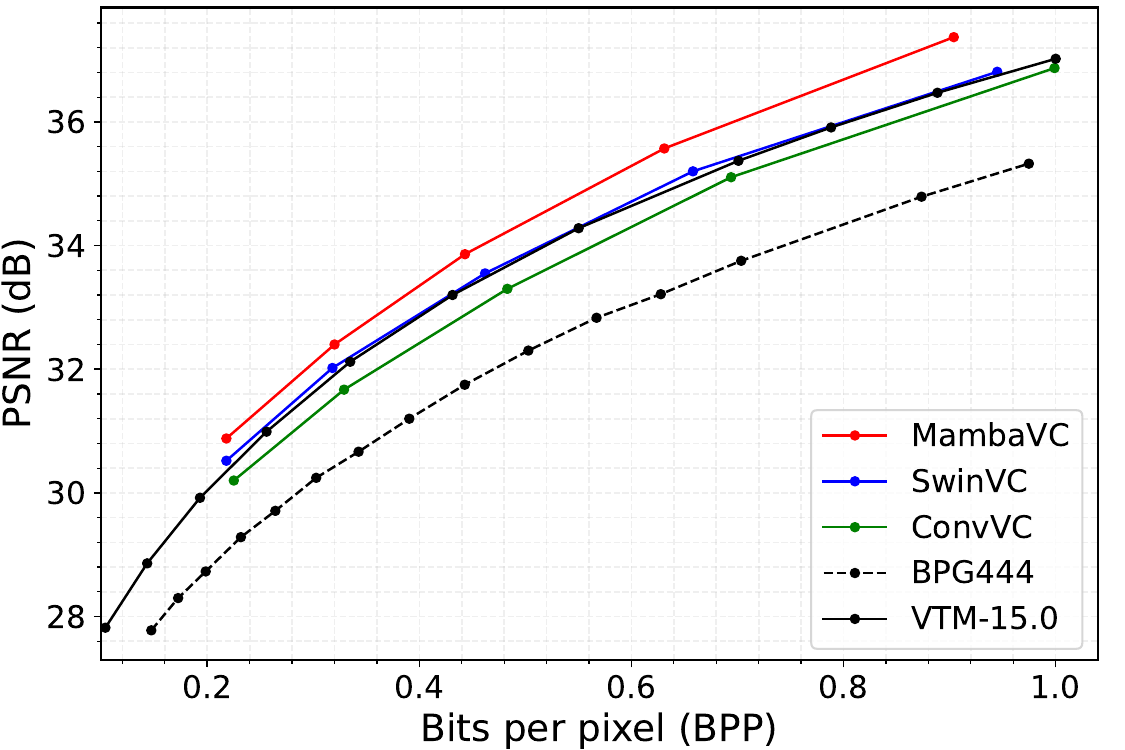}}
	\subfigure[Rate saving (\%) over VTM-15.0 (larger is better).] 
    {\includegraphics[width=.4965\textwidth]{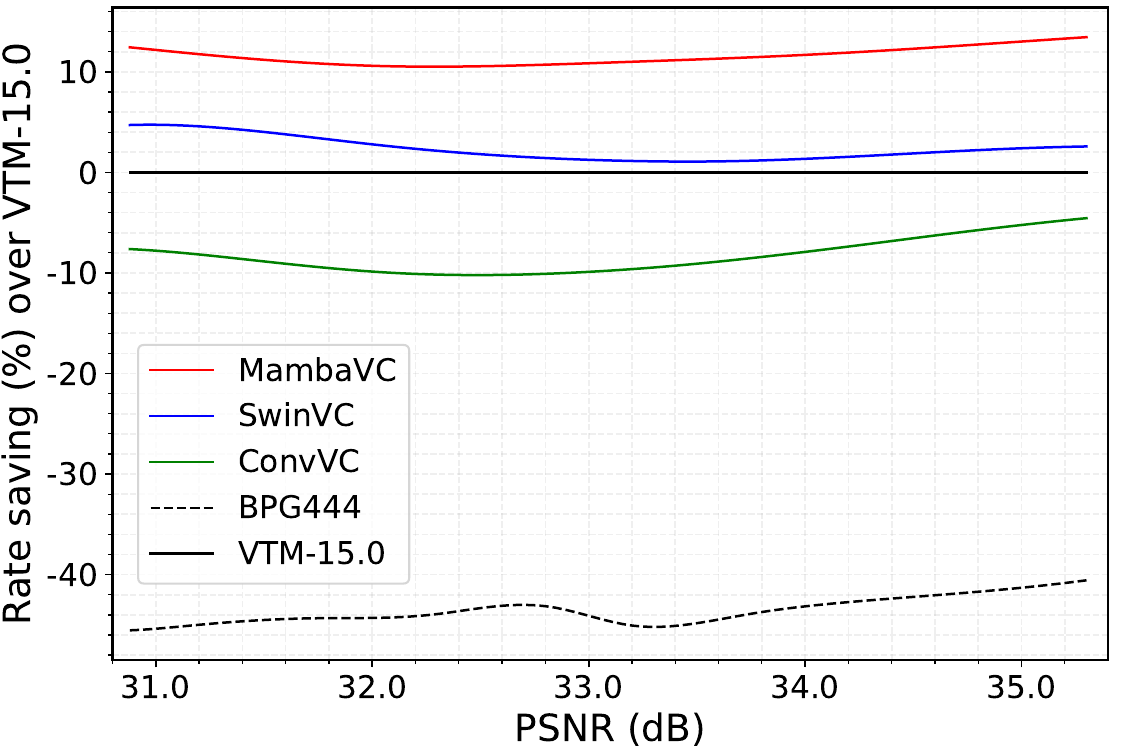}}
    \vspace{-0.3cm}
	\caption{Comparison of compression efficiency on JPEG-AI~\cite{jpegai} among different variants.}
	\label{JPEGAI_Var_performance}
\end{figure}

Additional rate-distortion results on Kodak~\cite{franzen1999kodak},CLIC2020~\cite{toderici2020clic} and JPEG-AI~\cite{jpegai} are shown in Figure~\ref{Var_performance}, Figure~\ref{CLIC_Var_performance} and Figure~\ref{JPEGAI_Var_performance}.
\clearpage

\begin{figure}[t]
    \centering
    \vspace{-0.3cm}
    \subfigure{  
        \begin{minipage}[b]{0.9\textwidth}  
            \includegraphics[width=1\textwidth]{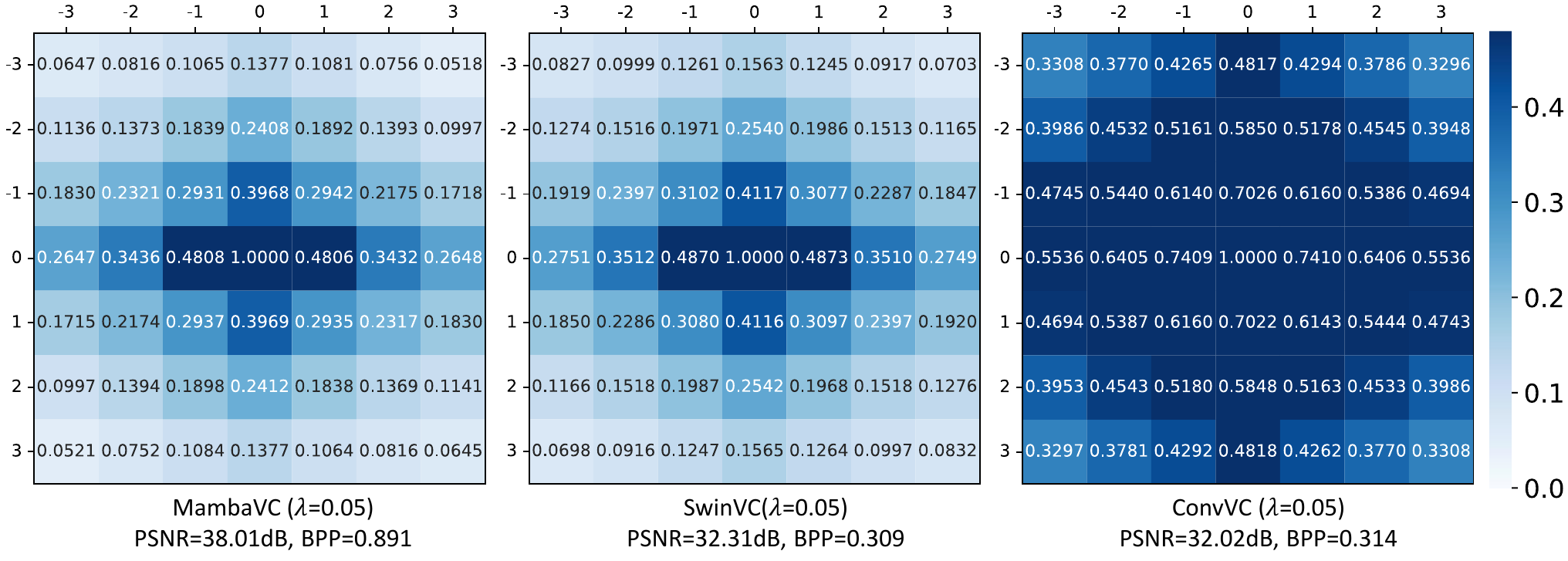} 
        \end{minipage}  
    } 
    \vspace{-0.3cm}
    \subfigure{  
        \begin{minipage}[b]{0.9\textwidth}  
            \includegraphics[width=1\textwidth]{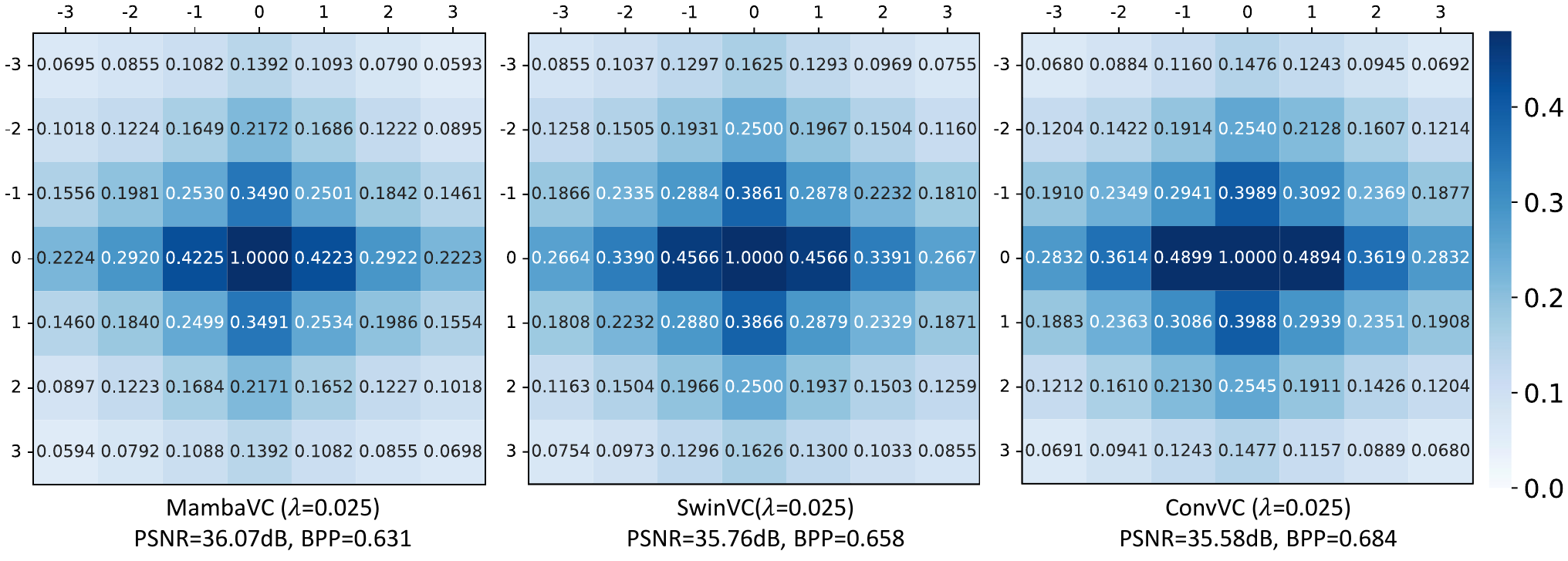} 
        \end{minipage}  
    }
    \vspace{-0.3cm}
    \subfigure{  
        \begin{minipage}[b]{0.9\textwidth}  \includegraphics[width=1\textwidth]{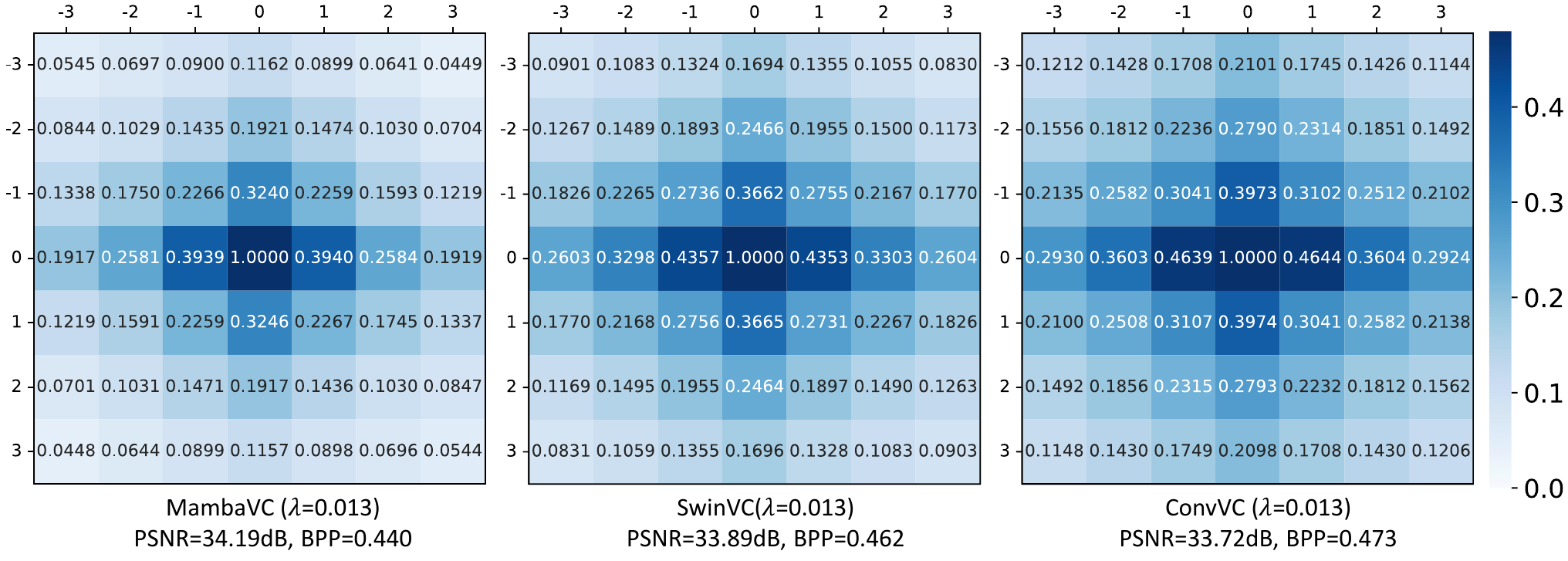} 
        \end{minipage}  
    }
    \vspace{-0.3cm}
    \subfigure{  
        \begin{minipage}[b]{0.9\textwidth}  
            \includegraphics[width=1\textwidth]{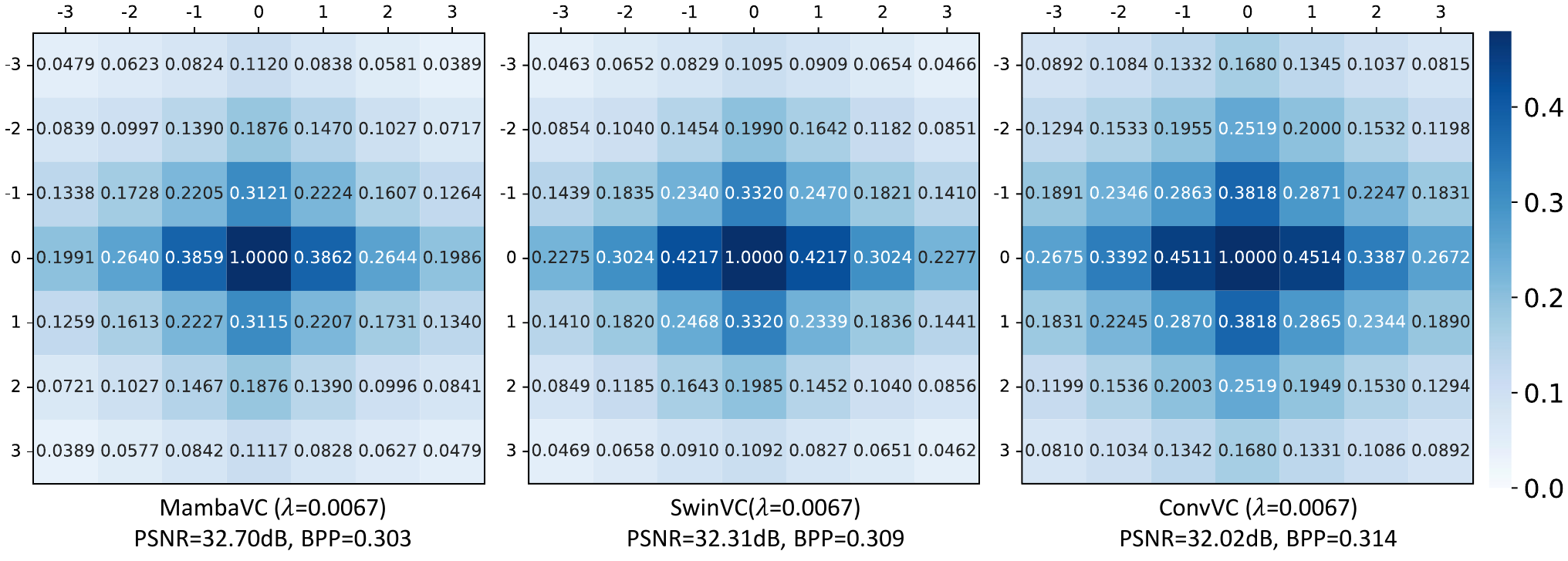} 
        \end{minipage}  
    } 
    \caption{Latent correlation of $(\bm{z}-\mu(\bm{z}))/\sigma(\bm{z})$, averaged across all latent elements of all images on Kodak~\cite{franzen1999kodak}. The value at position $(i,j)$ represents cross-correlation between spatial locations $(x,y)$ and $(x+i,y+j)$ along the channel dimension. Each row represents different variants trained with the same $\lambda$, with $\lambda$ values from top to bottom being 0.05, 0.025, 0.013, and 0.0067.}
    \label{Hyper Latent Correlation}
\end{figure}
\clearpage
\subsection{Hyper Latent Correlation}
Figure~\ref{Hyper Latent Correlation} illustrates the spatial correlation of the normalized prior latents. Horizontally comparing the different methods, MambaVC consistently shows the best performance across all $\lambda$. Vertically comparing the results, as the $\lambda$ decreases, the proportion of distortion loss diminishes, leading the model to focus more on compression ratio and thus eliminate more redundancy.

\begin{figure}[h]
    \vspace{-0.3cm}
	\centering
	\subfigure[\modelname{}] 
    {\includegraphics[width=.2\textwidth]{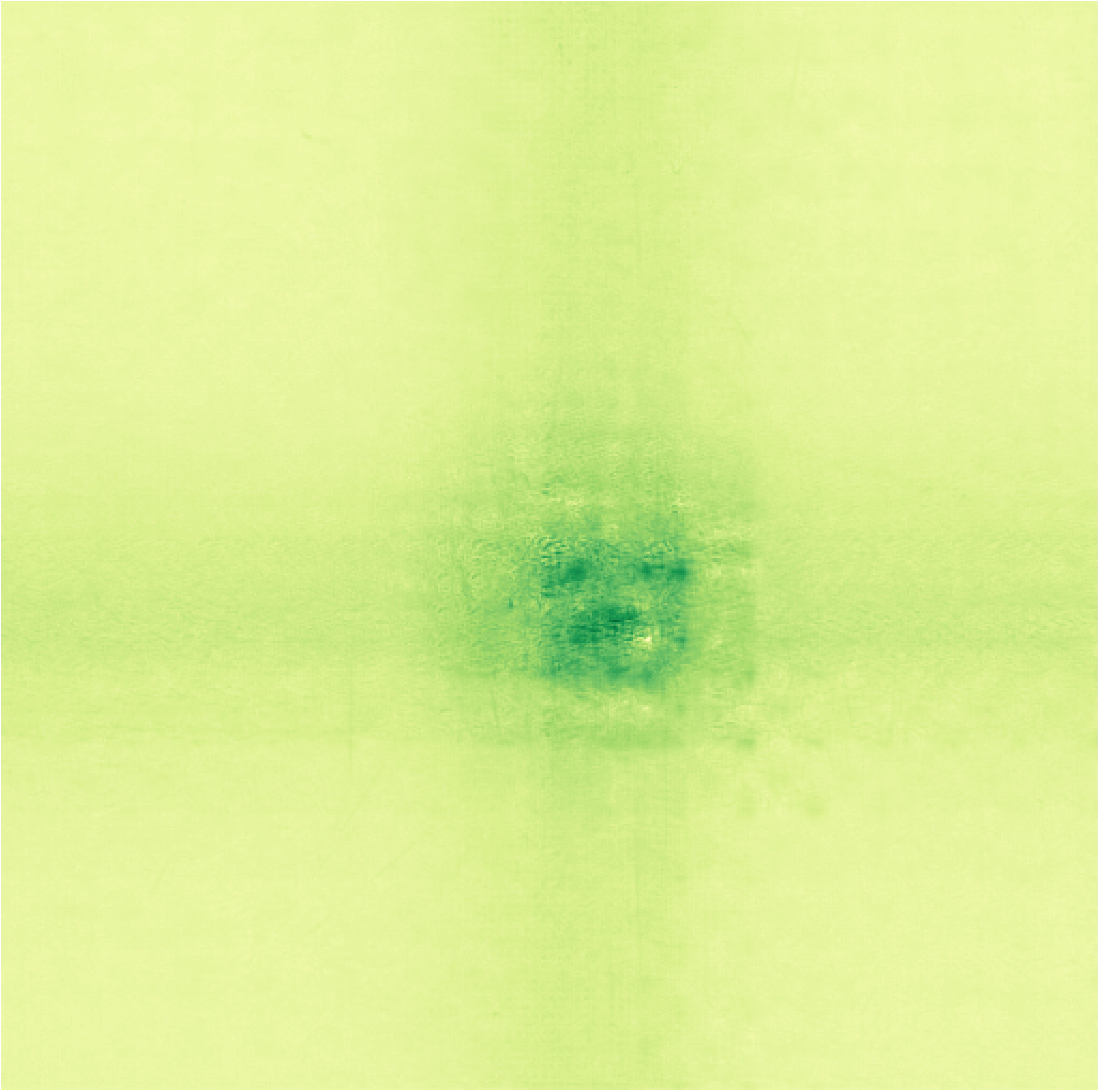}}
	\subfigure[SwinVC]     
    {\includegraphics[width=.2\textwidth]{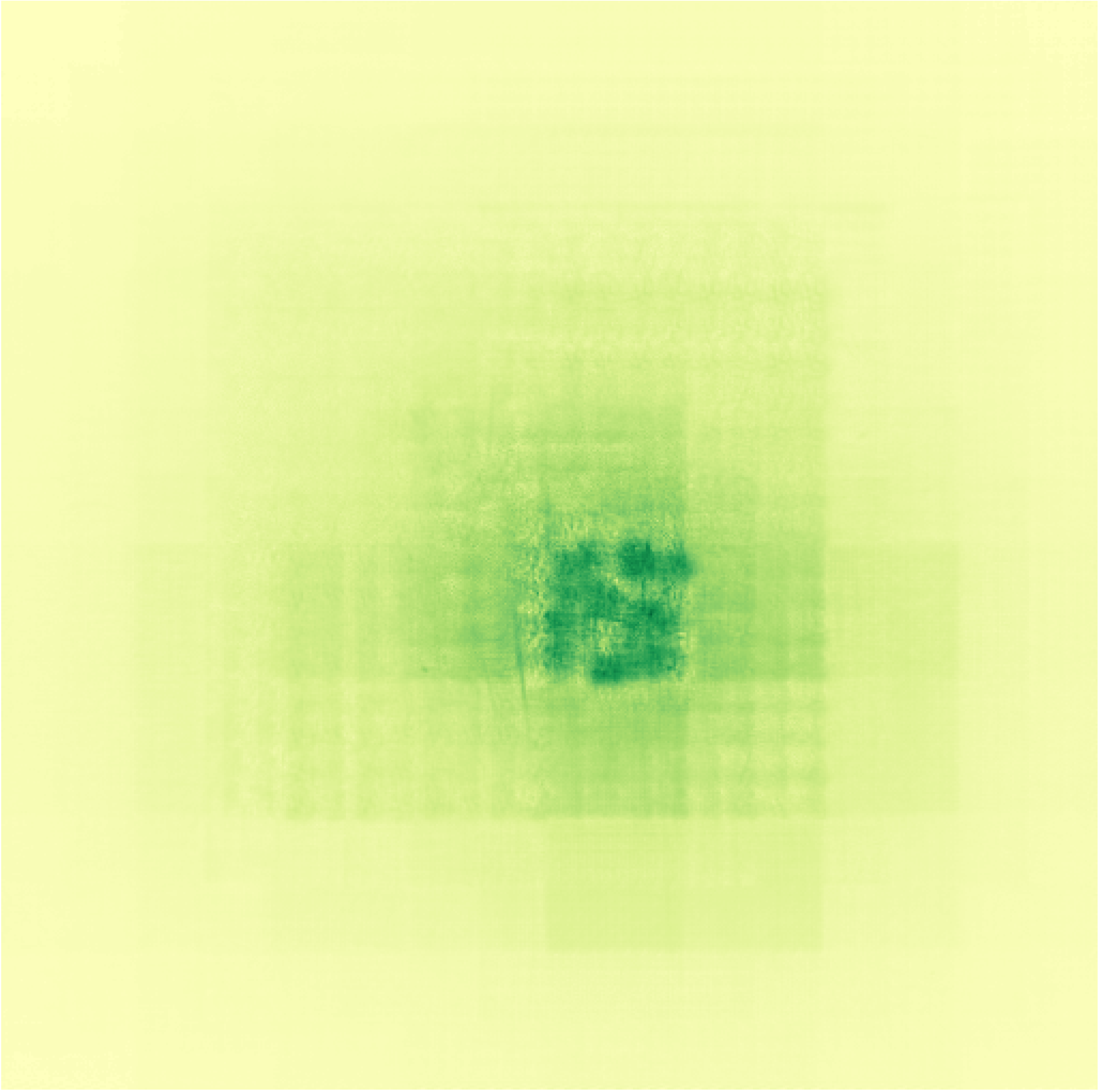}}
    \subfigure[ConvVC]     
    {\includegraphics[width=.2\textwidth]{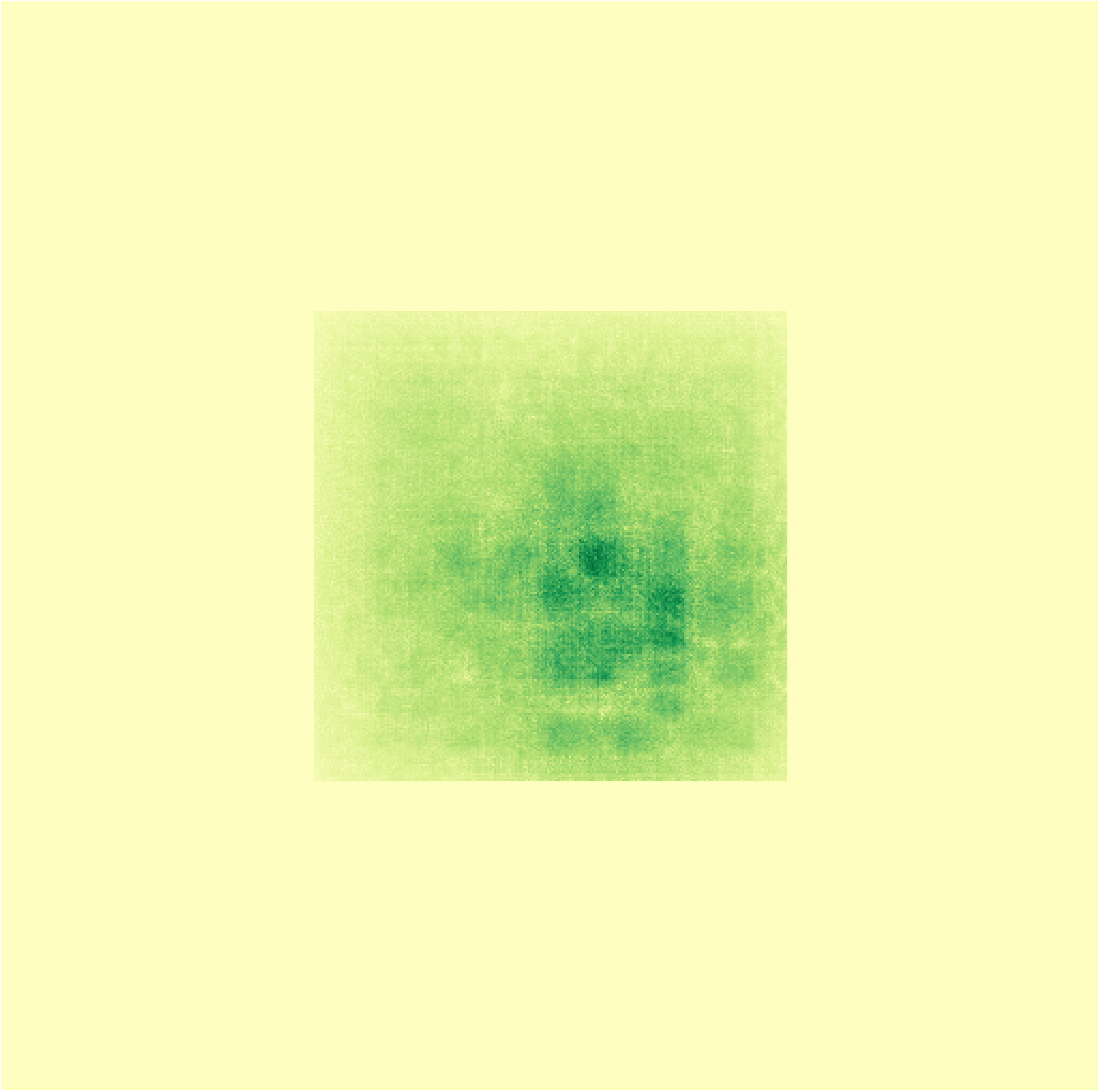}}
    \subfigure 
    {\includegraphics[width=.036\textwidth]{picture/ref.pdf}}
    \vspace{-0.1cm}
	\caption{Comparison of the ERF of the encoders and hyper encoder $g_a \circ h_a$ in \modelname{} and its variants on Kodak~\cite{franzen1999kodak}. Here, we calculate the absolute gradients $\left| \frac{d\bm{z}}{d\bm{x}} \right|$ of a pixel in the hyper latent $\bm{z}$.}
	\label{ERF_z}
    \vspace{-0.1cm}
\end{figure}
\subsection{Effective Receptive Field}
In Figure~\ref{ERF}, we present the receptive fields of latent $\bm{y} $after passing through the encoder $g_a$. Additionally, we explore the receptive fields of the hyper latent $\bm{z}$ after passing through the hyper encoder $g_a \circ h_a$, as shown in Figure~\ref{ERF_z}. Vertically comparing the methods, we observe that the receptive field expands as the network depth increases, suggesting a greater influence of surrounding areas on the value of each spatial point. Horizontally comparing the methods, MambaVC consistently demonstrates the largest receptive field among all approaches.

\end{document}